\documentclass[twocolumn,aps,pre,showpacs,preprint,amsmath,amssymb,showkeys,10pt]{revtex4-1} 

\usepackage[french,english]{babel}
\usepackage[T1]{fontenc}
\usepackage[utf8]{inputenc}
\usepackage{graphicx}
\usepackage{dsfont}
\usepackage{mathrsfs}
\usepackage{mathtools}
\usepackage[captionskip=-162pt]{subfig}
\usepackage{array}
\usepackage{hyperref}

\newcommand{\abs}[1]{\left\vert #1 \right\vert}
\newcommand{\mean}[1]{\left< #1 \right>}
\newcommand{\vect}[1]{\boldsymbol{#1}}
\newcommand{\mtx}[1]{\boldsymbol{#1}}
\newcommand{\Tr}{\textrm{Tr}}
\newcommand{\grad}{\vect{\nabla}}
\newcommand{\dv}{{\rm div}}
\newcommand{\etot}{E}
\newcommand{\ptot}{\vect{\mathcal{P}}}
\newcommand{\pj}{\mtx{P}}
\newcommand{\id}{\mtx{\mathrm{Id}}}
\newcommand{\rh}{\frac{r}{h}}
\newcommand{\rd}{{\rm d}}

\begin{document}

\title{Size consistency in Smoothed Dissipative Particle Dynamics}

\author{G\'er\^ome Faure}
\affiliation{ CEA, DAM, DIF, F-91297 Arpajon, France }
\author {Julien Roussel}
\affiliation{ Universit\'e Paris-Est, CERMICS (ENPC), INRIA, F-77455 Marne-la-Vall\'ee, France }
\author{Jean-Bernard Maillet}
\affiliation{ CEA, DAM, DIF, F-91297 Arpajon, France }
\author{Gabriel Stoltz}
\affiliation{ Universit\'e Paris-Est, CERMICS (ENPC), INRIA, F-77455 Marne-la-Vall\'ee, France }

\date{\today}

\begin{abstract}
  Smoothed Dissipative Particle Dynamics (SDPD) is a mesoscopic method which allows to select the level of resolution at which a fluid is simulated. In this work, we study the consistency of the resulting thermodynamic properties as a function of the size of the mesoparticles, both at equilibrium and out of equilibrium. We also propose a reformulation of the SDPD equations in terms of energy variables. This increases the similarities with Dissipative Particle Dynamics with Energy conservation and opens the way for a coupling between the two methods. Finally, we present a numerical scheme for SDPD that ensures the conservation of the invariants of the dynamics. Numerical simulations illustrate this approach.
\end{abstract}

\maketitle

\section{Introduction}
\label{sec:introduction}
With the development of new architectures and massively parallel codes, Molecular Dynamics (MD) simulations have been applied to systems of increasing sizes, up to billions of atoms, and times, up to a few nanoseconds~\cite{glosli_2007,kadau_2010}.
But MD simulations still cannot reach the time and length scales at which some complex phenomena, such as reactive waves in molecular systems, occur.
A variety of mesoscopic methods have been designed to stretch these scales by several orders of magnitude, with decreasing predictive power compared to MD.
They generally consider fewer degrees of freedom and allow for larger timesteps since they do not need to track the interatomic vibrations and use softer potentials.

Dissipative Particle Dynamics (DPD) has been introduced as a coarse-grained mesoscopic model~\cite{hoogerbrugge_1992} which represents groups of atoms, typically forming one or several molecules, by a single particle.
DPD particles interact through some potential energy (usually, a soft potential) while dissipative and stochastic forces are added to take into account the missing degrees of freedom.
The magnitude of the fluctuation and dissipation forces are related through a fluctuation-dissipation relation, ensuring the sampling of the canonical ensemble~\cite{espanol_1995}.

DPD can however only be used when the temperature in the system is fixed, which prevents its application to nonequilibrium situations such as shock waves.
DPD with conserved energy (DPDE)~\cite{avalos_1997,espanol_1997} has been introduced to cover such situations.
The coarse-grained internal degrees of freedom in DPD are represented in this model by a single variable, called internal energy, which exchanges energy with the external degrees of freedom through a dissipation and fluctuation mechanism.
This ensures the conservation of the total energy in the system and allows for its use in nonequilibrium situations such as the simulation of shock waves and reactive waves~\cite{stoltz_2006,maillet_2007,maillet_2011,strachan_2005,lin_2014,brennan_2014}.

However, DPDE has some shortcomings on its own.
While it is possible to choose the friction parameters without perturbing the equilibrium properties, as long as an appropriate fluctuation-dissipation relation is satisfied, it is still unclear whether it is possible to retrieve various dynamical properties such as the equilibration time between the internal and external degrees of freedom, and transport coefficients of the fluid (\emph{e.g.} self-diffusion, thermal conductivity, shear viscosity, \dots)~\cite{kroonblawd_2016}.
Moreover, the fluctuations in DPDE do not scale with the level of coarse-graining chosen to model the system and it is also questionable if one can satisfactorily represent several non-bonded particles by a single coarse-grained particle~\cite{bock_2007}.
Recent works such as~\cite{espanol_2016} aim at providing firm theoretical grounds for the replacement of a single molecule by a thermal blob obeying DPDE-like equations.
Such methods thus remains quite atomistic in their validity domains and do not allow to freely choose the level of resolution to be used.

On the other hand, Smoothed Particle Hydrodynamics (SPH)~\cite{lucy_1977,monaghan_1977} is a particular Lagrangian method to solve the Navier-Stokes equations.
It has become a popular method in astrophysics~\cite{monaghan_1977} and in meshless simulations of fluid dynamics~\cite{monaghan_1994,morris_1997}.
It is easy to implement since particles are used as interpolation nodes and no mesh is involved.
However it does not account for thermal fluctuations which can be of importance for phenomena such as hydrodynamic instabilities~\cite{kadau_2007} or for small fluid particles.

Smoothed Dissipative Particle Dynamics (SDPD)~\cite{espanol_2003} has been introduced to overcome the shortcomings of the previously mentioned methods.
It blends the SPH discretization of the Navier-Stokes equations with the thermal fluctuations of mesoscopic models, and thus allows to deal with hydrodynamics at the nanoscale.
SDPD has in particular been used to study colloids~\cite{vasquez_2009,bian_2012} or polymer suspensions~\cite{litvinov_2008}. 
It has been shown that SDPD yields a consistent diffusion coefficient for a colloid in a SDPD bath~\cite{vasquez_2009,litvinov_2009} at any resolution.

One of the main improvement of SDPD over DPDE is the ability to select the desired level of resolution in the model by playing with a parameter fixing the number of molecules one mesoparticle stands for.
This paves the way for multiscale simulations based on a concurrent coupling between models at different coarse-grained level, such as MD and SDPD~\cite{petsev_2015}, SDPD and Navier-Stokes~\cite{moreno_2013} or even SDPD at various resolutions~\cite{kulkarni_2013,petsev_2016}.
Our aim in this work is to justify the use of SDPD in a multiscale setting by showing that SDPD can reproduce the thermodynamic properties of a microscopic system at any resolution.
The method relies on fitting an equation of state obtained by MD simulations to convey the physical information needed in SDPD.
We will handle both equilibrium properties and dynamic properties in a nonequilibrium situation.
We also present a new numerical scheme which preserve the invariants of the system for the integration of the full SDPD dynamics involving the evolutions of the positions, momenta and internal energies.

This article is organized as follows.
We first present in Section~\ref{sec:equations} a reformulation of the original SDPD equations in terms of the internal energy rather than entropy to increase the structural similarity between SDPD and DPDE.
In Section~\ref{sec:schemes}, we introduce a numerical scheme inspired by the ones used for DPDE and study the energy conservation it provides.
Section~\ref{sec:results} is devoted to the numerical study of the size consistency.
We show that we are able to reproduce the microscopic equilibrium properties such as pressure or temperature for a wide range of resolutions in Section~\ref{sec:pg} and~\ref{sec:lj}.
We also study the effect of the SDPD resolution in the simulation of shock waves in Section~\ref{sec:shocks}.
Some conclusions are gathered in Section~\ref{sec:conclusion}, while more technical arguments are postponed to the Appendix.

\section{Smoothed Dissipative Particle Dynamics}
\label{sec:equations}
At the hydrodynamic scale, the dynamics of the fluid is governed by the Navier-Stokes equations~\eqref{eq:navier-stokes}, which read in their Lagrangian form when the heat conduction is neglected (for time $t\geq0$ and position $\vect{x}$ in a domain $\Omega\subset \mathbb{R}^3$):
\begin{equation}
  \label{eq:navier-stokes}
  \begin{aligned}
    {\rm D}_t\rho + \rho\,\dv_{\vect{x}}\vect{v} &= 0,\\
    \rho {\rm D}_t\vect{v} &= \dv_{\vect{x}}\left(\mtx{\sigma}\right),\\
    \rho{\rm D}_t\left(u + \frac12\vect{v}^2\right) &= \dv_{\vect{x}}\left(\mtx{\sigma}\vect{v}\right).
  \end{aligned}
\end{equation}
The material derivative used in the Lagrangian description is defined as
\[
  D_t f(t,\vect{x}) = \partial_t f(t,\vect{x}) + \vect{v}(t,\vect{x})\grad_{\vect{x}}f(t,\vect{x}).
\]
The unknowns are $\rho(t,\vect{x}) \in \mathbb{R}$ the density of the fluid, $\vect{v}(t,\vect{x}) \in \mathbb{R}^3$ its velocity, $u(t,\vect{x}) \in \mathbb{R}$ its internal energy and $\mtx{\sigma}(t,\vect{x}) \in \mathbb{R}^{3\times 3}$ the stress tensor:
\begin{equation}
\label{eq:stress-tensor}
  \mtx{\sigma} = P\id + \eta(\grad \vect{v} + (\grad\vect{v})^T) + \left(\zeta-\frac23\eta\right)\dv(\vect{v})\id,
\end{equation}
where $P$ is the pressure of the fluid, $\eta$ the shear viscosity and $\zeta$ the bulk viscosity.

In the following, we first present the SPH discretization of the Navier-Stokes equations in Section~\ref{sec:sph} before introducing the original SDPD equations proposed by Espa\~nol and Revenga~\cite{espanol_2003} in Section~\ref{sec:original-sdpd}.
We then formulate these equations in terms of internal energy variables in Section~\ref{sec:nrj-sdpd} by considering a fluctuation/dissipation mechanism structurally very similar to the one appearing in DPDE~\cite{avalos_1997,espanol_1997}. 
This reformulation in particular allows us to more easily construct stable and accurate numerical schemes which preserve the invariants of the system, such as the energy (see Section~\ref{sec:schemes}).
It also provides a first step towards a more straightforward coupling with DPDE although this issue is not discussed in this work.
In Section~\ref{sec:thermo}, we study the thermodynamic properties of the reformulated SDPD equations.

\subsection{Smoothed Particle Hydrodynamics}
\label{sec:sph}
Smoothed Particle Hydrodynamics~\cite{lucy_1977,monaghan_1977} is a Lagrangian discretization of the Navier-Stokes equations~(\ref{eq:navier-stokes}) on a finite number $N$ of fluid particles which play the role of interpolation nodes.
These fluid particles are associated with a portion of fluid of mass $m$.
They are located at positions $\vect{q}_i \in \Omega$ and have a momentum $\vect{p}_i \in\mathbb{R}^{3}$.
The internal degrees of freedom are represented by an entropy $S_i \in \mathbb{R}$.

\subsubsection{Approximation of field variables and their gradients}
\label{sec:approx-sph}

In the SPH discretization, the field variables are approximated as a weighted average of their values at the particle positions.
The weighting function $W$ is often referred to as a smoothing kernel function. 
The kernel function is generally chosen non negative, normalized as $\int_{\Omega} W(\vect{r})d\vect{r} = 1$, regular and with compact support~\cite{book_liu_2003,liu_2003}.
We denote by $h$ the smoothing length of the kernel $W$ so that $W(\vect{r})=0$ if $\abs{\vect{r}} \geq h $.
In the sequel, we use the notation $r = \abs{\vect{r}}$.
A variety of smoothing kernels have been introduced for the SPH discretization and compared in the litterature~\cite{fulk_1996,price_2012,liu_2003}.
A precise analysis of the kernel properties is however made difficult by the superposition of the two approximations made in the SPH discretization: the kernel approximation which consists in replacing the function $f$ by a smoothed version of it
\begin{equation}
  \label{eq:kernel-approx}
  f(x) \simeq \int_{\Omega}f(\vect{x'})W(\vect{x}-\vect{x'})d\vect{x'};
\end{equation}
and the particle approximation where the integral in equation~\eqref{eq:kernel-approx} is replaced by a summation over the particles.
The kernel approximation can be studied analytically and its accuracy improved by the choice of higher order kernels.
The particle approximation depends by nature on the particle configuration and can lead to undesired behaviors and instabilities such as particle clustering~\cite{monaghan_2000,swegle_1995,read_2010}.
Numerical simulations are thus required to evaluate the quality of the kernel.
In this work, since we do not focus on an exhaustive study of the influence of the kernel on the thermodynamic properties, we considered only two kernels: the Lucy function proposed early on~\cite{lucy_1977}:
\begin{equation}
  \label{eq:sdpd-lucy-w}
  W_{\rm Lucy}(\vect{r}) = \frac{105}{16\pi h^3}\left(1+3\frac{r}{h}\right)\left(1-\frac{r}{h}\right)^3 \mathds{1}_{r\leq h}.
\end{equation}
and, alternatively, a cubic spline~\cite{liu_2003}, whose expression reads
\begin{equation}
  \label{eq:sdpd-cubic-w}
  W_{\rm cubic}(\vect{r}) = \left\{
    \begin{array}{cl}
      \displaystyle \frac{8}{\pi h^3} \left(1-6\frac{r^2}{h^2}+6\frac{r^3}{h^3}\right) & \displaystyle \text{ if } r \leq \frac{h}{2},\\[1em]
      \displaystyle \frac{16}{\pi h^3} \left(1-\frac{r}{h}\right)^3 & \displaystyle \text{ if } \frac{h}{2} \leq r \leq h,\\[1em]
      0 & \displaystyle  \text{ if } r \geq h.
    \end{array}
  \right.
\end{equation}
The field variables are then approximated as
\begin{equation}
  \label{eq:sph-approx}
  f(\vect{x}) \approx \sum_{i=1}^N f_i W(\vect{x}-\vect{q}_i),
\end{equation}
where $f_i$ denotes the value of the field $f$ on the particle~$i$.

The approximation of the gradient $\grad_{\vect{x}} f$ is obtained by deriving equation~\eqref{eq:sph-approx}, which yields
\[
  \grad_{\vect{x}} f(\vect{x}) \approx  \sum_{i=1}^N f_i \grad_{\vect{x}}W(\abs{\vect{x}-\vect{q}_i}).
\]
In order to have more explicit expressions, we introduce the function $F$ such that $\vect{\nabla}_{\vect{r}} W(\vect{r}) = -F(\abs{\vect{r}})\vect{r}$.
For the Lucy function~(\ref{eq:sdpd-lucy-w}),
\[
F_{\rm Lucy}(r) = \frac{315}{4\pi h^5}\left(1-\frac{r}{h}\right)^2 \mathds{1}_{r \leq h},
\]
while, for the cubic spline~(\ref{eq:sdpd-cubic-w}),
\[
  F_{\rm cubic}(r) = \left\{
    \begin{array}{cl}
      \displaystyle \frac{48}{\pi h^5} \left(2-3\rh\right) & \displaystyle \text{ if } r \leq \frac{h}{2},\\[1em]
      \displaystyle \frac{48}{\pi h^5} \frac1{r} \left(1-\rh\right)^2& \displaystyle \text{ if } \frac{h}{2} \leq r \leq h,\\[1em]
      0 & \displaystyle \text{ if } r \geq h.
    \end{array}
    \right.
\]
The gradient approximation can then be rewritten as 
\[
  \grad_{\vect{x}} f(\vect{x}) \approx  -\sum_{i=1}^N f_i F(\abs{\vect{x}-\vect{q}_i})(\vect{x}-\vect{q}_i).
\]

In order to simplify the notation, we define the following quantities for two particles $i$ and $j$:
\[
  \vect{r}_{ij} = \vect{q}_i - \vect{q}_j,\quad
  r_{ij} = \abs{\vect{r}_{ij}},\quad
  \vect{e}_{ij} = \frac{\vect{r}_{ij}}{r_{ij}},\quad
  F_{ij} = F(r_{ij}).
\]
We can associate a density $\rho_i$ and volume $\mathcal{V}_i$ to each particle as
\begin{equation}
  \label{eq:sdpd-rho-v}
  \rho_i(\vect{q}) = \sum_{j=1}^N mW(\vect{r}_{ij}),\quad
  \mathcal{V}_i(\vect{q}) = \frac{m}{\rho_i(\vect{q})}.
\end{equation}
The corresponding approximations of the density gradient evaluated at the particle points read
\begin{equation}
  \label{eq:gradient-rho}
  \grad_{\vect{q}_j} \rho_i = \left\{
    \begin{array}{cl}
      m F_{ij}\vect{r}_{ij} & \text{ if } j\neq i,\\[.5em]
      -m \sum\limits_{j=1}^N F_{ij}\vect{r}_{ij} & \text{ if } j=i.
    \end{array}
  \right.
\end{equation}

\subsubsection{Thermodynamic closure}
\label{sec:thermo-closure}
As in the Navier-Stokes equations, an equation of state is required to close the set of equations provided by the SPH discretization.
This equation of state relates the internal energy $\varepsilon_i$ associated with particle $i$ with its density $\rho_i(\vect{q})$ (as defined by~\eqref{eq:sdpd-rho-v}) and its entropy $S_i$ through an internal energy function
\[
  \varepsilon_i(S_i,\vect{q}) = \mathcal{E}(S_i,\rho_i(\vect{q})).
\]
The equation of state $\mathcal{E}$ can be computed by microscopic simulations or by an analytic expression modeling the material behavior (see Section~\ref{sec:results} for some examples).
It is then possible to define, in accordance with the equation of state, a temperature
\[
  \mathcal{T}(S,\rho) = \partial_{S} \mathcal{E}(S,\rho),
\]
pressure
\[
  \mathcal{P}(S,\rho) = \frac{\rho^2}{m}\partial_{\rho}\mathcal{E}(S,\rho),
\]
and heat capacity at constant volume
\[
  \mathcal{C}(S,\rho) = \left[\frac{\partial_{S} \mathcal{E}}{\partial^2_{S}\mathcal{E}}\right](S,\rho).
\]
We assign to each particle the corresponding temperature $T_i$, pressure $P_i$ and heat capacity $C_i$ as
\[
\begin{aligned}
  T_i(S_i,\vect{q}) &= \mathcal{T}(S_i,\rho_i(\vect{q})),\\
  P_i(S_i,\vect{q}) &= \mathcal{P}(S_i,\rho_i(\vect{q})),\\
  C_i(S_i,\vect{q}) &= \mathcal{C}(S_i,\rho_i(\vect{q})).
\end{aligned}
\]
To simplify the notation, we omit in Sections~\ref{sec:eom-sph} and~\ref{sec:original-sdpd} the dependence of $T_i$, $P_i$ and $C_i$ on the variables $S_i$ and $\vect{q}$.

\subsubsection{Equations of motion}
\label{sec:eom-sph}
The SPH discretization can be split into two elementary dynamics, the first one being a conservative dynamics derived from the pressure gradient in the stress tensor~\eqref{eq:stress-tensor} and the last one a dissipative dynamics stemming from the viscous terms in~\eqref{eq:stress-tensor}.

The elementary force between particles $i$ and $j$ arising from the discretization of the pressure gradient in the Navier-Stokes momentum equation reads
\begin{equation}
  \label{eq:cons-forces}
  \vect{\mathcal{F}}_{{\rm cons},ij} = m^2\left(\frac{P_i}{\rho_i^2}+\frac{P_j}{\rho_j^2}\right)F_{ij}\vect{r}_{ij}.
\end{equation}
The corresponding total force is actually conservative since it can be rewritten in a Hamiltonian form as
\[
  \sum_{j\neq i}\vect{\mathcal{F}}_{{\rm cons},ij} = -\grad_{\vect{q}_{i}} H(\vect{q},\vect{p},S),
\]
where
\begin{equation}
  \label{eq:ham-entropy}
  H(\vect{q},\vect{p},S) = \sum_{i=1}^N \varepsilon_i(S_{i},\vect{q}) + \sum_{i=1}^N \frac{p_i^2}{2m}.
\end{equation}
This allows us to write the conservative part of the dynamics in Hamiltonian form as
\begin{equation}
  \label{eq:sdpd-cons}
  \left\{\begin{aligned}
      \rd\vect{q}_i &= \frac{\vect{p}_i}{m}\,\rd t,\\
      \rd\vect{p}_i &= \sum_{j\neq i} \vect{\mathcal{F}}_{{\rm cons},ij}\,\rd t,\\
      \rd S_i &= 0.
  \end{aligned}\right.
\end{equation}
This dynamics preserves by construction the total momentum $\displaystyle \sum_{i=1}^N \frac{\vect{p}_i^2}{2m}$ and the total energy $H(\vect{q},\vect{p},S)$.

In order to give the expression of the viscous part of the dynamics, we define the relative velocity for a pair of particles $i$ and $j$ as
\[
\vect{v}_{ij} = \frac{\vect{p}_i}{m}-\frac{\vect{p}_j}{m}.
\]
The viscous terms in the momentum continuity equation result in an elementary pairwise dissipative force
\[
\vect{\mathcal{F}}_{{\rm diss},ij} = - a_{ij}\vect{v}_{ij}-\left(\frac{a_{ij}}3+b_{ij}\right)(\vect{e}_{ij}\cdot\vect{v}_{ij})\vect{e}_{ij},
\]
where the friction coefficients are defined from the fluid viscosities $\eta$ and $\zeta$ appearing in the stress tensor~(\ref{eq:stress-tensor}) as
\[
a_{ij}=\left(\frac{5\eta}{3}-\zeta\right)\frac{m^2F_{ij}}{\rho_i\rho_j},\quad b_{ij}+\frac{a_{ij}}3 = 5\left(\frac{\eta}3+\zeta\right)\frac{m^2F_{ij}}{\rho_i\rho_j}.
\]
The pairwise dissipative elementary dynamics can be written as
\begin{equation}
  \label{eq:sdpd-diss-espanol}
  \left\{\begin{aligned}
      \rd\vect{q}_i &= \vect{0},\\
      \rd\vect{p}_i &= \sum_{j\neq i} \vect{\mathcal{F}}_{{\rm diss},ij}\,\rd t,\\
      T_i\rd S_i &= \frac12\sum_{j\neq i} \vect{v}_{ij}\cdot\vect{\mathcal{F}}_{{\rm diss},ij}\,\rd t.
  \end{aligned}\right.
\end{equation}
The third equation on the entropy is such that the total energy $H(\vect{q},\vect{p},S)$ is preserved.
In addition, Galilean invariance is ensured by the dynamics~(\ref{eq:sdpd-diss-espanol}) since $\sum\limits_{i=1}^N \rd\vect{p}_i = \vect{0}$.

\subsection{Original SDPD}
\label{sec:original-sdpd}

Smoothed Dissipative Particle Dynamics~\cite{espanol_2003} is a top-down mesoscopic method relying on the SPH discretization of the Navier-Stokes equations with the addition of thermal fluctuations which are modeled by a stochastic force.
SDPD is a set of stochastic differential equations for the variables used in the original SDPD equations: the positions $\vect{q}_i\in\Omega\subset\mathbb{R}^{3}$, the momenta $\vect{p}_i\in\mathbb{R}^{3}$ and the entropies $S_i\in \mathbb{R}$ for $i=1\dots N$.
The fluctuation terms read
\[
\rd\vect{\mathcal{F}}_{{\rm fluct},ij} = \left(A_{ij}\rd\mtx{\overline{W}}_{ij}+\frac13B_{ij}{\rm Tr}(\rd\mtx{W}_{ij})\id\right)\vect{e}_{ij},
\]
where the fluctuation amplitudes are given by
\begin{equation}
\label{eq:sdpd-fluct-coefficient}
A_{ij}^2 = 8k_{\rm{B}}a_{i,j}\frac{T_iT_j}{T_i+T_j},\quad
B_{ij}^2 = 12k_{\rm{B}}b_{i,j}\frac{T_iT_j}{T_i+T_j}.
\end{equation}
These coefficients are determined in~\cite{espanol_2003} through the GENERIC framework~\cite{grmela_1997}.
Here, for $1\leq i,j\leq N$, $\mtx{W}_{ij}$ are $3\times3$ matrices of independent standard Brownian motions such that $\mtx{W}_{ij}=-\mtx{W}_{ji}$.
We denote by $ \displaystyle \mtx{\overline{W}}_{ij} = \frac12\left(\mtx{W}_{ij}+\mtx{W}_{ij}^T\right) - \frac13\Tr\left(\mtx{W}_{ij}\mtx{\rm Id}\right)$ the symmetric traceless part of $\mtx{W}_{ij}$.
The random fluctuation term is balanced by an extra dissipative force
\[
\vect{\widetilde{\mathcal{F}}}_{{\rm diss},ij} = d_{ij}\vect{\mathcal{F}}_{{\rm diss},ij},
\]
with
\[
d_{ij} = k_{\rm B}\frac{T_iT_j}{(T_i+T_j)^2}\left(\frac1{C_i}+\frac1{C_j}\right).
\]

The dissipative and stochastic forces produce a variation of the entropy which is accounted for by a dynamics on the entropy $S_i$ in order to ensure the conservation of the energy $H(\vect{q},\vect{p},S)$.
The elementary fluctuation/dissipation dynamics reads
\begin{equation}
  \label{eq:sdpd-fluct-espanol}
  \left\{\begin{aligned}
      \rd\vect{q}_i =&\, \vect{0},\\
      \rd\vect{p}_i =& \sum_{j\neq i} (1 - d_{ij})\vect{\mathcal{F}}_{{\rm diss},ij}\,\rd t + \rd\vect{\mathcal{F}}_{{\rm fluct},{ij}},\\
      T_i\rd S_i =& \sum\limits_{j\neq i} \frac12\mathscr{D}_{ij} \vect{v}_{ij}^T\vect{\mathcal{F}}_{{\rm diss},ij}\,\rd t\\
      &- \frac{8k_{\rm B}}{m}\frac{T_iT_j}{T_i+T_j}\left(\frac{10}3a_{ij}+b_{ij}\right)\rd t\\
      &-\frac12 \vect{v}_{ij}^T\rd\vect{\mathcal{F}}_{{\rm fluct},ij},
    \end{aligned}\right.
\end{equation}
where $\mathscr{D}_{ij} = 1-d_{ij} + \frac{T_i}{T_i+T_j}\frac{k_{\rm B}}{C_i}$.
The dynamics~(\ref{eq:sdpd-fluct-espanol}) replaces the dissipative dynamics~(\ref{eq:sdpd-diss-espanol}) of the SPH discretization.
It can be shown that it preserves the energy $H(\vect{q},\vect{p},S)$ and the total momentum.

Finally, the complete set of equations of motion for the original SDPD~\cite{espanol_2003} is obtained by concatenating the conservative dynamics~(\ref{eq:sdpd-cons}) and the fluctuation/dissipation dynamics~(\ref{eq:sdpd-fluct-espanol}) as
\begin{equation}
  \label{eq:sdpd-espanol}
  \left\{
  \begin{aligned}
    \rd\vect{q}_i =&\, \vect{v}_i\,\rd t,\\
    \rd\vect{p}_i =& \sum_{j\neq i} m^2\left(\frac{P_i}{\rho_i^2}+\frac{P_j}{\rho_j^2}\right)F_{ij}\vect{r}_{ij}\,\rd t\\
    &- (1-d_{ij})\left[a_{ij}\vect{v}_{ij}+\left(\frac{a_{ij}}3+b_{ij}\right)(\vect{v}_{ij}^T\vect{e}_{ij})\vect{e}_{ij}\right]\rd t \\
    &+ \left(A_{ij}\rd\mtx{\overline{W}}_{ij}+\frac13B_{ij}{\rm Tr}(\rd\mtx{W}_{ij})\id\right)\vect{e}_{ij},\\
    T_i\rd S_i =& \sum_{j\neq i} \frac12\mathscr{D}_{ij}\left(a_{ij}\vect{v}_{ij}^2+\left(\frac{a_{ij}}3+b_{ij}\right)(\vect{v}_{ij}^T\vect{e}_{ij})^2\right)\rd t\\
    &-\frac{8k_{\rm B}}{m}\frac{T_iT_j}{T_i+T_j}\left(\frac{10}3a_{ij}+b_{ij}\right)\rd t\\
    &-\frac12 \vect{v}_{ij}^T\left(A_{ij}\rd\mtx{\overline{W}}_{ij}+\frac13B_{ij}{\rm Tr}(\rd\mtx{W}_{ij})\id\right)\vect{e}_{ij}.
  \end{aligned}
  \right.
\end{equation}
The dynamics~\eqref{eq:sdpd-espanol} preserve the total momentum $\sum\limits_{i=1}^N\vect{p}_i$ and the total energy $H(\vect{q},\vect{p},S)$ since the elementary dynamics~(\ref{eq:sdpd-cons}) and~(\ref{eq:sdpd-fluct-espanol}) conserve these invariants.
The GENERIC framework~\cite{grmela_1997} ensures that measures of the form
\begin{equation}
  \label{eq:sdpd-espanol-minv}
  \begin{aligned}
    &\nu(\rd\vect{q}\,\rd\vect{p}\,\rd S)\\
    &\,= g\left(H(\vect{q},\vect{p},S),\sum\limits_{i=1}^N\vect{p}_i\right)\prod_{i=1}^N\frac{\exp\left(\frac{S_i}{k_{\rm B}}\right)}{T_i(S_i,\vect{q})}\,\rd\vect{q}\,\rd\vect{p}\,\rd S
  \end{aligned}
\end{equation}
are invariant for the dynamics~\eqref{eq:sdpd-espanol}.

\subsection{Energy reformulation}
\label{sec:nrj-sdpd}

We propose in this section a reformulation of the original SDPD equations~(\ref{eq:sdpd-espanol}) in terms of positions, momenta and energies $\varepsilon_i\geq0$.
The corresponding phase space is denoted by $\mathscr{E} = \Omega^N\times \mathbb{R}^{3N}\times \mathbb{R}_+^N$.
We also propose a somewhat simpler expression for the fluctuation term in analogy with the stochastic term used in DPDE~\cite{avalos_1997,espanol_1997}.
The new expressions for the fluctuation/dissipation forces are chosen to ensure the same invariant measure as in the original SDPD equations.
Moreover, the same friction forces appear in both formulations.
The interest of this reformulation is twofold: first, it allows for a better control of the energy conservation in the integration scheme, second, it is a first step towards a more straightforward coupling with DPDE.

As in Section~\ref{sec:sph}, we need an input equation of state in order to close the equations.
In the energy formulation, the equation of state links the entropy $S_i$ with the internal energy $\varepsilon_i$ and the density $\rho_i(\vect{q})$ of the particles as
\begin{equation}
  \label{eq:sdpd-eos}
  S_i(\varepsilon_i,\vect{q})=\mathcal{S}(\varepsilon_i,\rho_i(\vect{q})).
\end{equation}
The pressure, temperature and heat capacity are then determined accordingly as
\begin{equation}
  \label{eq:eos-thermo}
  \begin{aligned}
    \mathcal{T}(\varepsilon,\rho) &= \left[\frac1{\partial_{\varepsilon}\mathcal{S}}\right](\varepsilon,\rho),\\
    \mathcal{P}(\varepsilon,\rho) &= -\frac{\rho^2}{m}\left[\frac{\partial_{\rho}\mathcal{S}}{\partial_{\varepsilon}\mathcal{S}}\right](\varepsilon,\rho),\\
    \mathcal{C}(\varepsilon,\rho) &= -\left[\frac{(\partial_{\varepsilon} \mathcal{S})^2}{\partial_{\varepsilon}^2\mathcal{S}}\right](\varepsilon,\rho),
  \end{aligned}
\end{equation}
and assigned to each particle as
\[
\begin{aligned}
  T_i(\varepsilon_i,\vect{q}) &= \mathcal{T}(\varepsilon_i,\rho_i(\vect{q})),\\
  P_i(\varepsilon_i,\vect{q}) &= \mathcal{P}(\varepsilon_i,\rho_i(\vect{q})),\\
  C_i(\varepsilon_i,\vect{q}) &= \mathcal{C}(\varepsilon_i,\rho_i(\vect{q})).
\end{aligned}
\]
As in the previous sections, we now omit the dependence of $T_i$, $P_i$ and $C_i$ on the variables $\varepsilon_i$ and $\vect{q}$.

We keep the conservative part of the dynamics~(\ref{eq:sdpd-cons}).
In order to reformulate it in terms of internal energies, we compute the associated variation in the energy, expressed in the new set of independent variables $(\vect{q},\vect{p},\varepsilon)$ as
\[
  E(\vect{q},\vect{p},\varepsilon) = \sum_{i=1}^N \varepsilon_i + \frac{\vect{p}_i^2}{2m}.
\]
Since a change of variables should not change the energy, $H$ defined by~\eqref{eq:ham-entropy} and $E$ are related by
\[
  E(\vect{q},\vect{p},\varepsilon) = H \Big(\vect{q},\vect{p},\mathcal{S}(\varepsilon_1,\rho_i(\vect{q})),\dots,\mathcal{S}(\varepsilon_N,\rho_i(\vect{q}))\Big).
\]
The dynamics~(\ref{eq:sdpd-cons}) being isentropic, the induced energy variation is simply given by
\begin{equation}
  \label{eq:energy-variation}
  \rd \varepsilon_i = - P_i\rd\mathcal{V}_i.
\end{equation}
From the definition~(\ref{eq:sdpd-rho-v}) of the density and volume of a particle, the infinitesimal volume variation reads $\displaystyle \rd\mathcal{V}_i = -\frac{m}{\rho_i^2}\rd\rho_i$ with $\rd\rho_i$ given by
\[
\rd\rho_i = -\sum_{j\neq i} mF_{ij}\vect{r}_{ij}\cdot\vect{v}_{ij}\,\rd t.
\]
Therefore, the conservative part of the dynamics can be rewritten as
\begin{equation}
  \label{eq:sdpd-nrj-cons}
  \left\{\begin{aligned}
      \rd\vect{q}_i &= \frac{\vect{p}_i}{m}\,\rd t,\\
      \rd\vect{p}_i &= \sum_{j\neq i} \vect{\mathcal{F}}_{{\rm cons},ij}\,\rd t,\\
      \rd\varepsilon_i &= -\sum_{j\neq i}\frac{m^2P_i}{\rho_i(\vect{q})^2}F_{ij}\vect{r}_{ij}\cdot\vect{v}_{ij}\,\rd t.
  \end{aligned}\right.
\end{equation}
Let us emphasize that the indices $i$ and $j$ do not play a symmetrical role in the evolutions of the internal energies, although the conservative forces are symmetric.
This asymmetry is a consequence of the fact that the energy variation~(\ref{eq:energy-variation}) for a given particle only involves the volume variation of this particle.

In the spirit of DPDE, we choose a pairwise fluctuation and dissipation term for $i<j$ of the following form
\begin{equation}
  \label{eq:sdpd-simple-fluct}
  \left\{
  \begin{aligned}
    \rd\vect{p}_i &= -\mtx{\Gamma}_{ij}\vect{v}_{ij}\,\rd t + \mtx{\Sigma}_{ij}\rd\vect{B}_{ij},\\
    \rd\vect{p}_j &= \mtx{\Gamma}_{ij}\vect{v}_{ij}\,\rd t - \mtx{\Sigma}_{ij}\rd\vect{B}_{ij},\\
    \rd\varepsilon_i &= \frac12\left[\vect{v}_{ij}^T\mtx{\Gamma}_{ij}\vect{v}_{ij} - \frac{\Tr(\mtx{\Sigma}_{ij}\mtx{\Sigma}_{ij}^T)}{m}\right]\rd t -\frac12 \vect{v}_{ij}^T\mtx{\Sigma}_{ij}\rd\vect{B}_{ij},\\
    \rd\varepsilon_j &= \frac12\left[\vect{v}_{ij}^T\mtx{\Gamma}_{ij}\vect{v}_{ij} - \frac{\Tr(\mtx{\Sigma}_{ij}\mtx{\Sigma}_{ij}^T)}{m}\right]\rd t -\frac12 \vect{v}_{ij}^T\mtx{\Sigma}_{ij}\rd\vect{B}_{ij},
  \end{aligned}
  \right.
\end{equation}
where $\vect{B}_{ij}$ is a $3$-dimensional vector of standard Brownian motions, $\mtx{\Gamma}_{ij}$ and $\mtx{\Sigma}_{ij}$ are $3\times3$ symmetric matrices.
In the dynamics~\eqref{eq:sdpd-simple-fluct}, the equations acting on the momenta preserve the total momentum in the system.
Furthermore, as in DPDE, the equations for the energy variables are determined to ensure the conservation of the total energy $E(\vect{q},\vect{p},\varepsilon)$.
As $\displaystyle \rd \varepsilon_i = -\frac12 \rd \left(\frac{\vect{p}_i^2}{2m} + \frac{\vect{p}_j^2}{2m}\right)$, It\^o calculus yields the resulting equations in~\eqref{eq:sdpd-simple-fluct}.

For the friction and fluctuation coefficients, we consider matrices of the form
\begin{equation}
  \label{eq:fluct-gamma}  
  \mtx{\Gamma}_{ij} = \gamma^{\parallel}_{ij}\pj^{\parallel}_{ij} + \gamma^{\perp}_{ij}\pj^{\perp}_{ij},\quad 
  \mtx{\Sigma}_{ij} = \sigma^{\parallel}_{ij}\pj^{\parallel}_{ij} + \sigma^{\perp}_{ij}\pj^{\perp}_{ij},
\end{equation}
with the projection matrices $\pj^{\parallel}_{ij}$ and $\pj^{\perp}_{ij}$ given by
\[
  \pj_{ij}^{\parallel} = \vect{e}_{ij}\otimes\vect{e}_{ij},\quad
  \pj_{ij}^{\perp} = \id - \pj_{ij}^{\parallel}
\]

Using such a decomposition for the friction and fluctuation matrices, the dynamics~(\ref{eq:sdpd-simple-fluct}) is obtained by superposing the following dynamics for $\theta \in \{\parallel, \perp\}$:
\begin{equation}
  \label{eq:sdpd-fd-elem}
  \left\{
  \begin{aligned}
    \rd\vect{p}_i =& -\gamma_{ij}^{\theta}\pj_{ij}^{\theta}\vect{v}_{ij}\rd t + \sigma_{ij}^{\theta}\pj_{ij}^{\theta}\rd\vect{B}_{ij},\\
    \rd\vect{p}_j =&\, \gamma_{ij}^{\theta}\pj_{ij}^{\theta}\vect{v}_{ij}\rd t - \sigma_{ij}^{\theta}\pj_{ij}^{\theta}\rd\vect{B}_{ij},\\
    \rd\varepsilon_i =&\, \frac12\left[\gamma_{ij}^{\theta}\vect{v}_{ij}^T\pj_{ij}^{\theta}\vect{v}_{ij} - \frac{(\sigma_{ij}^{\theta})^2}{m}\Tr(\pj_{ij}^{\theta})\right]\rd t\\
    &-\frac12 \sigma_{ij}^{\theta}\vect{v}_{ij}^T\pj_{ij}^{\theta}\rd\vect{B}_{ij},\\
    \rd\varepsilon_j =&\, \frac12\left[\gamma_{ij}^{\theta}\vect{v}_{ij}^T\pj^{\theta}\vect{v}_{ij} - \frac{(\sigma_{ij}^{\theta})^2}{m}\Tr(\pj_{ij}^{\theta})\right]\rd t\\
    &-\frac12 \sigma_{ij}^{\theta}\vect{v}_{ij}^T\pj_{ij}^{\theta}\rd\vect{B}_{ij}.
  \end{aligned}
  \right.
\end{equation}
The choice
\begin{equation}
  \label{eq:sdpd-gamma-sigma}
  \begin{aligned}
    \gamma_{ij}^{\parallel} &= \left(\frac43a_{ij}+b_{ij}\right) \left( 1 - d_{ij}\right),\\
    \gamma_{ij}^{\perp} &= a_{ij}\left( 1 - d_{ij} \right),\\
    \sigma_{ij}^{\theta} &= 2\sqrt{\frac{\gamma_{\theta}}{1-d_{ij}} k_{\rm B}\frac{T_iT_j}{T_i+T_j}},
  \end{aligned}
\end{equation}
with the coefficients $a_{ij}$ and $b_{ij}$ defined in Section~\ref{sec:eom-sph}, ensures that measures of the form
\begin{equation}
  \label{eq:sdpd-energy-minv}
  \begin{aligned}
    &\mu(\rd\vect{q}\,\rd\vect{p}\,\rd \varepsilon)\\
    &\,= g\left(E(\vect{q},\vect{p},\varepsilon),\sum\limits_{i=1}^N\vect{p}_i\right)\prod_{i=1}^N\frac{\exp\left(\frac{S_i(\varepsilon_i,\vect{q})}{k_{\rm B}}\right)}{T_i(\varepsilon_i,\vect{q})}\,\rd\vect{q}\,\rd\vect{p}\,\rd \varepsilon
  \end{aligned}
\end{equation}
are left invariant by the elementary dynamics~\eqref{eq:sdpd-simple-fluct}  (see Appendix~\ref{sec:app-minv} for the proof).
Note that $\mu$ is just obtained from the measure $\nu$ defined in~\eqref{eq:sdpd-espanol-minv} by the change of variables $(\vect{q},\vect{p},S) \to (\vect{q},\vect{p},\varepsilon)$.
As our derivation shows, other choices are possible for the coefficients $\gamma_{ij}^{\theta}$ and $\sigma_{ij}^{\theta}$ (see~\eqref{eq:generic-gamma} in Section~\ref{sec:minv-fd}).
This may be of interest since it is possible to choose a constant fluctuation magnitude $\sigma_{ij}^{\theta}$ while the friction coefficient $\gamma_{ij}^{\theta}$ still depends on the configuration of the system through the positions $\vect{q}$ and the internal energies $\varepsilon_i$ and $\varepsilon_j$ as
\[
  \gamma_{ij}^{\theta} = \frac14 \left(T_iT_j(\partial_{\varepsilon_i}+\partial_{\varepsilon_j})\left[\frac{(\sigma_{ij}^{\theta})^2}{T_iT_j}\right] +  \frac{(\sigma_{ij}^{\theta})^2}{k_{\rm B}} \frac{T_i+T_j}{T_iT_j} \right).
\]
Such a choice would further increase the similarity with DPDE and simplify the numerical discretization.
However, in this work, we stick to the choice~(\ref{eq:sdpd-gamma-sigma}) which yields the same friction terms as in the original SDPD equations~\eqref{eq:sdpd-espanol}.

As a result, the SDPD equations of motion reformulated in the position, momentum and internal energy variables read
\begin{equation}
  \label{eq:sdpd-energy}
  \left\{
  \begin{aligned}
    \rd\vect{q}_i =&\, \frac{\vect{p}_i}{m}\,\rd t,\\
    \rd\vect{p}_i =& \sum_{j\neq i} m^2\left(\frac{P_i}{\rho_i^2}+\frac{P_j}{\rho_j^2}\right)F_{ij}\vect{r}_{ij}\,\rd t - \mtx{\Gamma}_{ij}\vect{v}_{ij}\,\rd t\\
    &+ \mtx{\Sigma}_{ij}\rd\vect{B}_{ij},\\
    \rd \varepsilon_i =& \sum_{j\neq i} -\frac{m^2P_i}{\rho_i^2}F_{ij}\vect{r}_{ij}^T\vect{v}_{ij}\,\rd t\\
    &+ \frac12 \left[\vect{v}_{ij}^T\mtx{\Sigma}_{ij}\vect{v}_{ij} -\frac1{m}\Tr(\mtx{\Sigma}_{ij}\mtx{\Sigma}_{ij}^T)\right]\rd t\\
    & - \frac12 \vect{v}_{ij}^T\mtx{\Sigma}_{ij}\rd\vect{B}_{ij},
  \end{aligned}
  \right.
\end{equation}
with $\mtx{\Sigma}_{ij}$ and $\mtx{\Gamma}_{ij}$ given by~\eqref{eq:fluct-gamma} and~(\ref{eq:sdpd-gamma-sigma}).
The dynamics~\eqref{eq:sdpd-energy} preserves the total momentum $\sum\limits_{i=1}^N\vect{p}_i$ and the total energy $E(\vect{q},\vect{p},\varepsilon)$ since all the elementary sub-dynamics ensure these conservations (see Appendix~\ref{sec:app-cons} for the proof).
Let us also emphasize that the reformulated dynamics involves only a $3$-dimensional Brownian motion $\vect{B}_{ij}$ for each pair instead of the $6$-dimensional Brownian motion $\mtx{\overline{W}}_{ij}$ appearing in the original dynamics~\eqref{eq:sdpd-fluct-espanol}.

\subsection{Thermodynamic properties of the reformulated SDPD}
\label{sec:thermo}

We present in this section expressions for the estimators of thermodynamic quantities like temperature and pressure.
Following the same ideas as for DPDE~\cite{homman_2016}, we rely on a thermodynamic equivalence with an appropriate canonical measure to make the computations tractable.

Although the dynamics~(\ref{eq:sdpd-energy}) leaves any measure of the form~\eqref{eq:sdpd-energy-minv} invariant (as shown in Appendix~\ref{sec:app-minv}), there are no mathematical results about its ergodicity since the fluctuation may be degenerate.
Even for DPD, ergodicity is known to hold only for simple one-dimensional systems~\cite{shardlow_2006}.
Since the total energy and the total momentum are conserved, we assume the ergodicity of the dynamics~\eqref{eq:sdpd-energy} with respect to the measure
\[
  \begin{aligned}
    &\mu_{E_0,\ptot_0}(\rd\vect{q}\rd\vect{p}\rd \varepsilon) \\
    & \,= Z^{-1}_{E_0,\ptot_0}\delta\left(E(\vect{q},\vect{p},\varepsilon)-E_0\right)\delta\left(\sum\limits_{i=1}^N\vect{p}_i -\ptot_0\right)\\
    &\,\phantom{=}\,\times\prod_{i=1}^N \frac{\exp\left(\frac{S_i(\varepsilon_i,\vect{q} )}{k_{\rm B}}\right)}{T_i(\varepsilon_i,\vect{q})}\,\rd\vect{q}\,\rd\vect{p}\,\rd \varepsilon,
  \end{aligned}
\]
with $E_0$ the initial total energy, $\ptot_0$ the initial momentum and $Z^{-1}_{E_0,\ptot_0}$ a normalization constant.
Under this assumption, the average of some observable $A$ can be estimated as
\[
\mean{A}_{E_0,\ptot_0} = \int_{\mathcal{E}} A\,\rd\mu_{E_0,\vect{\mathcal{P}}_0} = \underset{t\to+\infty}{\rm lim} \frac1{t} \int_0^t A(q_s,p_s,\varepsilon_s)\rd s,
\]
where $(q_s,p_s,\varepsilon_s)$ is the solution at time $s$ of~\eqref{eq:sdpd-energy}.
We assume in the following that $\ptot_0 = \vect{0}$.
This can be achieved by adopting the center of mass reference frame.

To justify the expressions of thermodynamic estimators of temperature, it is convenient to introduce the canonical measure
\begin{equation}
  \label{eq:sdpd-minveq}
  \begin{aligned}
    &\mu_{\beta}(\rd\vect{q}\,\rd\vect{p}\,\rd\varepsilon) \\
    &\,= Z^{-1}_{\beta}\prod_{i=1}^N\frac{\exp\left(-\beta\left[\frac{\vect{p}_i^2}{2m} + \varepsilon_i\right]+\frac{S_i(\varepsilon_i,\vect{q})}{k_{\rm B}}\right)}{T_i(\varepsilon_i,\vect{q})}\,\rd\vect{q}\,\rd\vect{p}\,\rd \varepsilon,
  \end{aligned}
\end{equation}
where $\beta$ is chosen such that $\mean{E}_{\mu_{\beta}} = E_0$ and $Z^{-1}_{\beta}$ a normalization constant.
In the thermodynamic limit, $\mu_{\beta}$ and $\mu_{E_0,\vect{0}}$ are expected to be equivalent in the same way that the microcanonical and canonical measures are equivalent for systems described only in terms of $\vect{q}$ and $\vect{p}$.

Under the canonical measure~(\ref{eq:sdpd-minveq}), the thermodynamic temperature $\displaystyle T_{\beta} = \frac1{k_{\rm B}\beta}$ can be estimated from the kinetic energy as
\[
  T_{\beta} = \mean{\frac{\vect{p}_i^2}{3mk_{\rm B}}}_{\mu_{\beta}},
\]
which motivates the use of the kinetic temperature
\begin{equation}
\label{eq:estim-tkin}
  T_{\rm kin}(\vect{p}) = \frac1{N}\sum\limits_{i=1}^N\frac{\vect{p}_i^2}{3mk_{\rm B}}
\end{equation}
as an estimator of $T_{\beta}$.
Under some assumptions on the equation of state~\eqref{eq:sdpd-eos}, which hold for instance for the ideal gas equation of state (see Equation~\eqref{eq:pg-eos} below), namely
\begin{equation}
  \label{eq:entropy-assumptions}
  \begin{aligned}
    \forall \rho \in \mathbb{R}_+, &\quad \mathcal{S}(\rho,\varepsilon)\xrightarrow[\varepsilon\to 0]{}-\infty,\\
    &\quad \mathcal{S}(\rho,\varepsilon) - k_{\rm B}\beta\varepsilon\xrightarrow[\varepsilon\to+\infty]{}-\infty,
  \end{aligned}
\end{equation}
the internal temperature also provides an estimator of the thermodynamic temperature since
\begin{equation}
  \label{eq:estim-tint}
  T_{\rm \beta} = \mean{T_i}_{\mu_{\beta}}.
\end{equation}
The temperature $T_{\beta}$ can therefore be estimated from the average of the internal temperature in the system as
\[
  T_{\rm int} = \frac1N \sum\limits_{i=1}^N T_i.
\]
Let us stress that the internal temperature estimator for SDPD relies on an arithmetic average in contrast to DPDE where an harmonic mean should be used~\cite{stoltz_2006}.

The thermodynamic pressure in the system is defined as the derivative of the free energy $\mathcal{F}$ with respect to the total volume $\mathcal{V} = \abs{\Omega}$ of the system.
With the previous ergodicity assumption, the pressure can be estimated as
\begin{equation}
  \label{eq:estim-pressure}
  P = -\partial_{\mathcal{V}} \mathcal{F} = P_{\rm kin} + P_{\rm virial},
\end{equation}
where $P_{\rm kin}$ is the kinetic pressure
\[
P_{\rm kin} = \frac{N}{\mathcal{V}\beta},
\]
and $P_{\rm virial}$ is the virial pressure
\[
P_{\rm virial} = \frac1{3\mathcal{V}}\mean{\sum_{1\leq i<j \leq N}\vect{\mathcal{F}}_{{\rm cons},ij}\cdot\vect{r}_{ij}}_{\mu_{\beta}}.
\]
A detailed proof of the equalities~(\ref{eq:estim-tint}) and~(\ref{eq:estim-pressure}) can be read in Appendix~\ref{sec:app-tp}.

\subsection{Scaling properties of SDPD}
\label{sec:scaling-sdpd}
One of the important feature of SDPD is that it is possible to prescribe a size for the particles, which enables a multiscale approach~\cite{vasquez_2009,kulkarni_2013,petsev_2016}.
In the following, we study the behavior of SDPD when the mass of the particles varies (see Section~\ref{sec:results}).
In this perspective, the mass of the fluid particles is given by $m_K = K m_0$, where $m_0$ is the mass of one microscopic particle (\emph{i.e.} a molecule).
It is in fact convenient to define a system of reduced units for each size $K$:
\begin{equation}
  \label{eq:reduced-units}
  \begin{aligned}
    \widetilde{m}_K &= m_K,\\
    \widetilde{l}_K &= \left(\frac{m_K}{\rho}\right)^{\frac13},\\
    \widetilde{\varepsilon}_K &= K k_{\rm B}T,
  \end{aligned}
\end{equation}
where $\widetilde{m}_K$ is the mass unit, $\widetilde{l}_K$ the length unit, $\widetilde{\varepsilon}_K$ the energy unit and $\rho$ the average density of the fluid.
With such a set of reduced units, the time unit is 
\[
  \widetilde{t}_K = \widetilde{l}_K\sqrt{\frac{\widetilde{m}_K}{\widetilde{\varepsilon}_K}} = \frac{m_0^{\frac12}K^{\frac13}}{\rho^{\frac13} \sqrt{k_{\rm B}T}}.
\]

The smoothing length $h_K$ defining the cut-off radius in~\eqref{eq:sdpd-lucy-w} and~\eqref{eq:sdpd-cubic-w} also needs to be adapted to the size of the SDPD particles so that the approximations~(\ref{eq:sph-approx}) continue to make sense.
In order to keep the average number of neighbors roughly constant in the smoothing sum, $h_K$ should be rescaled as
\[
  h_K = h\left(\frac{m_K}{\rho}\right)^{\frac13}.
\]
In this work, we have taken $h=2.5$, which correspond to a typical number of 60-70 neighbors, a commonly accepted number~\cite{liu_2003}.

\section{Numerical scheme}
\label{sec:schemes}
To our knowledge, there are very few works providing numerical schemes for the integration of the full energy-conserving SDPD such as Gatsonis \emph{et al.}~\cite{gatsonis_2014} who mix a Velocity-Verlet scheme for the update of positions and momenta and a Runge-Kutta scheme for the entropy updates.
In such works, no specific attention is devoted to the preservation of the invariants such as the energy.
Let us also mention that most of the work published for SDPD resort to a simplified version of the dynamics where internal temperatures are kept fixed at $T_i=T_{\rm ref}$.
The equations of motion are then integrated using a Verlet scheme or more specific splitting schemes designed for SDPD~\cite{litvinov_2010}.
These schemes are very much inspired by integration schemes designed for DPD such as the Shardlow splitting scheme~\cite{shardlow_2003}.

When considering the dynamics~(\ref{eq:sdpd-espanol}) or~(\ref{eq:sdpd-energy}), the preservation of the invariants, especially the energy, requires some care in the design of the numerical scheme.
Other desirable properties include stability, accuracy and parallelizability.
Exhibiting a scheme satisfying all these constraints is not an easy task.
There is, to our knowledge, no numerical scheme for SDPD able to meet these requirements.
Though the development of such an integration scheme is not the purpose of this work, we suggest in Section~\ref{sec:splitting} a convenient scheme inspired by works done for DPDE~\cite{homman_2016}.
The increased similarity of the reformulated dynamics~(\ref{eq:sdpd-energy}) with DPDE makes it indeed possible to resort to similar integration schemes in both methods.
We analyze the properties of the scheme in terms of energy conservation in Section~\ref{sec:vnum}.
A more detailed comparison with existing schemes (such as~\cite{gatsonis_2014}) and adaptions of other DPDE schemes to the SDPD setting is currently in progress.

\subsection{Splitting scheme for SDPD}
\label{sec:splitting}

We propose a numerical scheme to integrate the stochastic dynamics~(\ref{eq:sdpd-energy}) obtained from the superposition of the elementary dynamics~(\ref{eq:sdpd-nrj-cons}) and~(\ref{eq:sdpd-simple-fluct}).
We denote by $\Delta t$ the time step.
Since the SDPD and DPDE equations have a similar structure, we follow the ideas introduced for the discretization of the DPD equations~\cite{shardlow_2003} and their adaptation to the DPDE model~\cite{stoltz_2006,homman_2016}.
The corresponding schemes are splitting schemes, which are a popular method to integrate differential equations (as proposed in~\cite{trotter_1959,strang_1968}) and stochastic differential equations.
The scheme presented in the following is based on a Trotter splitting of the dynamics~\eqref{eq:sdpd-energy}.
First the conservative dynamics~\eqref{eq:sdpd-nrj-cons} is integrated with a Velocity-Verlet scheme (see~\eqref{eq:sdpd-verlet} below) during a time $\Delta t$, then the fluctuation/dissipation part~\eqref{eq:sdpd-simple-fluct} is approximately evolved during a time $\Delta t$ by successive pairwise updates (see~\eqref{eq:sdpd-shardlow} below).

We first consider the conservative dynamics~\eqref{eq:sdpd-nrj-cons} formulated in terms of entropies as~\eqref{eq:sdpd-cons}.
Since it is of Hamiltonian nature, it is natural to solve this part of the dynamics with a symplectic scheme such as the well-known Velocity Verlet scheme~\cite{verlet_1967} which is able to conserve the energy in average in the long term~\cite{hairer_2006}:
\begin{equation}
  \label{eq:sdpd-verlet}
  \left\{
  \begin{aligned}
    \vect{\tilde{p}}^{n+\frac12}_i &= \vect{p}^n_i + \sum_{j\neq i}\vect{\mathcal{F}}_{{\rm cons},ij}^n \frac{\Delta t}{2},\\
    \vect{q}^{n+1}_i &= \vect{q}^n_i + \frac{\vect{\tilde{p}}^{n+\frac12}}{m}\Delta t,\\
    \vect{\tilde{p}}^{n+1}_i &= \vect{\tilde{p}}^{n+\frac12}_i + \sum_{j\neq i}\vect{\mathcal{F}}_{{\rm cons},ij}^{n+1}\frac{\Delta t}{2}.
  \end{aligned}
  \right.
\end{equation}
Since the entropy $S_i$ of each particle is preserved by the integration of this reversible dynamics, the internal energy $\varepsilon_i$ after this step can be computed by inverting the equation of state as
\[
  \tilde{\varepsilon}_i^{n+1} = \mathcal{E}\left(S_i^n,\rho_i(\vect{q}^{n+1})\right),
\]
with the updated density $\rho_i(\vect{q}^{n+1})$.
If no analytic form is available for $\mathcal{E}$, a numerical inversion may be required (see Section~\ref{sec:lj} for more details).

To deal with the fluctuation/dissipation part, we use  a modification of the Shardlow scheme~\cite{shardlow_2003} which relies on a splitting of the fluctuation/dissipation into pairwise elementary dynamics~\eqref{eq:sdpd-simple-fluct}, each pair being handled successively.
If we consider the internal energies to be constant, the equations on the momenta become an Ornstein-Uhlenbeck process that can be solved analytically.
As suggested by Marsh~\cite{marsh_1998}, and later used in~\cite{stoltz_2006}, it is possible to ensure the conservation of the energy by a redistribution of the kinetic energy variation induced by the dissipative and stochastic forces in the internal energies.
We define the kinetic energy of the pair of particles $i$ and $j$ as
\[
E_{\rm kin}(\vect{p}_i,\vect{p}_j) = \frac{\vect{p}_i^2}{2m} + \frac{\vect{p}_j^2}{2m}.
\]
We also introduce, for $\theta \in \{\parallel,\perp\}$
\[
  \alpha_{ij}^{\theta} = \exp\left(-\frac{2\gamma_{ij}^{\theta}\Delta t}{m}\right),\quad
  \zeta_{ij}^{\perp} = \sigma_{ij}^{\theta}\sqrt{\frac{m(1-(\alpha_{ij}^{\theta})^2) }{4\gamma_{ij}^{\theta}}}.
\]

For a random Gaussian vector $\vect{G}_{ij}$, we define the updated momenta as
\[
\begin{aligned}
  &\vect{\Pi}_{ij}(\vect{p}_i,\vect{p}_j,\varepsilon_i,\varepsilon_j,\vect{G}_{ij}) = \begin{pmatrix}
    \vect{p}_i\\[.5em]
    \vect{p}_j
  \end{pmatrix}\\
  &\quad +  \sum\limits_{\theta\in\{\parallel,\perp\}} \pj_{ij}^{\theta}\left[\frac{m}2  (\alpha_{ij}^{\theta}-1) \vect{v}_{ij} + \zeta_{ij}^{\theta} \vect{G}_{ij} \right]
  \begin{pmatrix}
    \vect{1}\\[.5em]
    \vect{-1}
  \end{pmatrix},
\end{aligned}
\]
and the induced kinetic energy variation when changing $(\vect{p}_i,\vect{p}_j)$ to $\vect{\Pi}_{ij}(\vect{p}_i,\vect{p}_j,\varepsilon_i,\varepsilon_j,\vect{G}_{ij})$ as
\[
  \begin{aligned}
    &\Delta_{\vect{\Pi}}E_{\rm kin}(\vect{p}_i,\vect{p}_j,\varepsilon_i,\varepsilon_j,\vect{G}_{ij}) =\\
    &\quad E_{\rm kin} ( \vect{\Pi}_{ij}(\vect{p}_i,\vect{p}_j,\varepsilon_i,\varepsilon_j,\vect{G}_{ij})) - E_{\rm kin}(\vect{p}_i,\vect{p}_j).
  \end{aligned}
\]
We would like to stress that $\gamma^{\theta}$ and $\sigma^{\theta}$ include cut-off functions in their expressions, which limits the range of the fluctuation/dissipation interactions and allows an evaluation of the corresponding forces scaling linearly with the system size.
The integration of the elementary fluctuation/dissipation dynamics~\eqref{eq:sdpd-simple-fluct} then proceeds as follows.
We denote by $\vect{\overline{p}}^{n,ij}$ and $\overline{\varepsilon}^{n,ij}$ the momentum and internal energy at the moment just before the dynamics for the pair $(i,j)$ is integrated: $\vect{\overline{p}}_k^{n,ij}$ and $\overline{\varepsilon}_k^{n,ij}$ have thus been obtained by performing a step of the Verlet scheme~\eqref{eq:sdpd-verlet} and by integrating the previous pairs.
We generate a vector of 3 standard Gaussian variables $\vect{G}_{ij}^n$.
The integration of the pair $(i,j)$ consists in replacing $(\vect{\overline{p}}_i^{n,ij},\vect{\overline{p}}_j^{n,ij},\overline{\varepsilon}_i^{n,ij},\overline{\varepsilon}_j^{n,ij})$ by
\begin{equation}
  \label{eq:sdpd-shardlow}
    \begin{pmatrix}
      \displaystyle \vect{\Pi}_{ij}(\vect{\overline{p}}_i^{n,ij},\vect{\overline{p}}_j^{n,ij},\overline{\varepsilon}_i^{n,ij},\overline{\varepsilon}_j^{n,ij},\vect{G}_{ij}^n)\\[.5em]
      \displaystyle \overline{\varepsilon}_i^{n,ij} + \frac12\Delta_{\vect{\Pi}}E_{\rm kin}(\vect{\overline{p}}_i^{n,ij},\vect{\overline{p}}_j^{n,ij},\overline{\varepsilon}_i^{n,ij},\overline{\varepsilon}_j^{n,ij},\vect{G}_{ij}^n)\\[.5em]
      \displaystyle \overline{\varepsilon}_j^{n,ij} + \frac12\Delta_{\vect{\Pi}}E_{\rm kin}(\vect{\overline{p}}_i^{n,ij},\vect{\overline{p}}_j^{n,ij},\overline{\varepsilon}_i^{n,ij},\overline{\varepsilon}_j^{n,ij},\vect{G}_{ij}^n)
    \end{pmatrix}.
\end{equation}
As a consequence of the energy redistribution, the total energy is exactly conserved by these elementary updates.

The final scheme for the integration of the reformulated dynamics consists in the superposition of the Verlet scheme~\eqref{eq:sdpd-verlet} and of the Shardlow-like scheme~\eqref{eq:sdpd-shardlow} for all pairs $1\leq i<j\leq N$.

This scheme is essentially sequential as it requires considering each pair of particles one after another.
The availability of parallel integration schemes is obviously crucial to apply the method to large physical systems.
It is still a challenge to design an accurate and efficient parallel algorithm for this kind of stochastic dynamics.
Some recent results were obtained for DPDE with a parallelization of Shardlow-like algorithms~\cite{larentzos_2014} or with the Splitting with Energy Reinjection integration (SER) which allows for an easy parallelization at the cost of a larger energy drift~\cite{homman_2016}.
The SER scheme has been introduced for DPDE but the similarities between the DPDE and SDPD equations make it possible to adapt SER for the dynamics~\eqref{eq:sdpd-energy}.

The two parts of the scheme separately ensure a good energy conservation: the Velocity-Verlet scheme~\eqref{eq:sdpd-verlet} preserves energy in average while the algorithm~\eqref{eq:sdpd-shardlow} preserves it exactly.
However, the overall scheme obtained by superposing~\eqref{eq:sdpd-verlet} and~\eqref{eq:sdpd-shardlow} can lead to energy drifts as observed in DPDE~\cite{lisal_2011,homman_2016} (see Section~\ref{sec:vnum}).
A standard numerical analysis shows that the scheme is of weak order 1 so that the error on average properties is of order $\Delta t$~\cite{leimkuhler_2015,homman_2016,homman_2016_phd}.

\subsection{Analysis of the energy drift}
\label{sec:vnum}
To validate our numerical scheme, we consider the ideal gas, whose
equation of state is given by
\begin{equation}
  \label{eq:pg-eos}
  \mathcal{S}_{\rm ideal}(\varepsilon,\rho) = \frac32(K-1)k_{\rm B}\ln(\varepsilon) - \frac12(K-1)\ln(\rho),
\end{equation}
where $K$ is the size of the SDPD particles.
We recall that this equation of state satisfies the conditions~\eqref{eq:entropy-assumptions}.
The equation of state is formulated in the reduced units introduced in Section~\ref{sec:scaling-sdpd}.

\begin{figure}[!ht]
  \centering
  \includegraphics{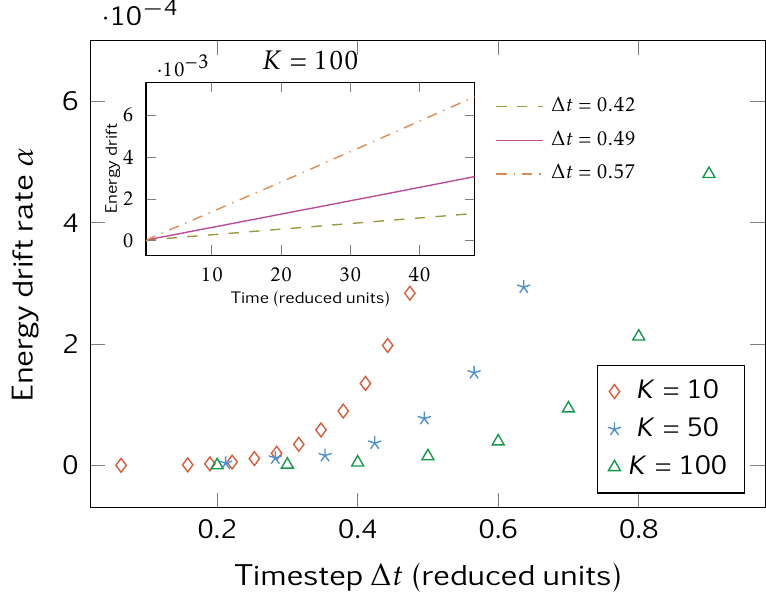}
  \caption{Average energy drift rate $\alpha_{\Delta t,K}$ as a function of the time step $\Delta t$ for the integration of SDPD with the scheme presented in Section~\ref{sec:splitting}. The inset represents the average time-dependent energy drift for $K=100$.}
  \label{fig:sdpd-energy-drift}
\end{figure}

We first check the energy conservation for different particle sizes $K$.
The scheme presented in Section~\ref{sec:splitting} displays a linear drift in energy with respect to the simulation time (see the inset in Figure~\ref{fig:sdpd-energy-drift}).
The results have been obtained by averaging over $n_{\rm sim} = 1000$ realizations of the dynamics for a total time $\tau_s = 50$ for each time step $\Delta t$ and size $K$.
We choose a unitary mass $m_0=10^{-25}$~kg.
The system is initialized as follows:
the particles are initially located on a simple cubic lattice at density $\rho = 1150$~kg.m$^{-3}$.
Their velocities are distributed according to a normal distribution with variance corresponding to a temperature $T = 1000$~K.
The internal energies are chosen so that $T_i(\varepsilon_i,\rho_i(\vect{q})) = T$ with the density $\rho_i(\vect{q})$ evaluated from the initial distribution of the positions.
Unless otherwise specified, all the numerical results presented in this work are obtained by resorting to this initialization method.
Also, in all the simulations of this work, the bulk viscosity is neglected while the shear viscosity is chosen as $\eta = 2\cdot10^{-3}$~Pa.s.

To analyze the error in the energy conservation, we compute, for each timestep $\Delta t$ and size $K$, the drift rate $\alpha_{\Delta t,K}$ defined as the slope of the energy drift.
We evaluate $\alpha_{\Delta t,K}$ by fitting a linear function on the energy with a least-square minimization.
The drift rate $\alpha_{\Delta t,K}$ is represented in Figure~\ref{fig:sdpd-energy-drift} as a function of $\Delta t$ for $K=10$, $K=50$ and $K=100$.
Assuming that the drift $\alpha_{\Delta t,K}$ can be written as
\[
  \alpha_{\Delta t,K} = \mathcal{C}_{K} \Delta t^{n_{K}},
\]
we find, by performing a least-square fit in a log-log scale, that $n_K$ is independent of $K$, with $n_K \approx 5.44$.
The prefactor $\mathcal{C}_K$ varies with the mass of the particle.
Here we estimate $\mathcal{C}_{10} = 9.50$, $\mathcal{C}_{50} = 181$ and $\mathcal{C}_{100} = 217$.

This suggests a way to choose the timestep to obtain a given drift rate: we perform a preliminary run with some timestep and measure the drift rate in this simulation.
The power law, independent of $K$, then allows us to estimate the timestep needed to keep the energy drift below a given threshold.
The results of Figure~\ref{fig:sdpd-energy-drift} also show that, for larger particle sizes $K$, we can increase the timestep in reduced units for a given drift rate, which allows us to integrate over longer times when using a coarse resolution.

\section{Study of the size consistency}
\label{sec:results}
In this section, we study how the properties predicted by SDPD are influenced by the choice of the resolution, namely the size of the particles.
We first study the size consistency of SDPD at equilibrium for both the ideal gas equation of state~\eqref{eq:pg-eos} (see Section~\ref{sec:pg}) and an equation of state optimized for a Lennard-Jones fluid~\cite{johnson_1993} (see Section~\ref{sec:lj}).
We then consider nonequilibrium situations with the simulation of shock waves in Section~\ref{sec:shocks}.

\subsection{Ideal gas}
\label{sec:pg}
We first study a SDPD system with the ideal gas equation of state~\eqref{eq:pg-eos} at different masses $m_K = K m_0$ with $m_0=10^{-25}$~kg.
We run simulations of a system of $N = 1000$ particles initialized according to the method described in Section~\ref{sec:vnum} at density $\rho =1150$~kg.m$^{-3}$ and temperature $T=1000$~K.
The number of iterations is fixed to $N_{\rm it}=5\times10^5$.
The timestep is chosen for each $K$ such that the drift in relative energy is less than $0.5 \%$ after $N_{\rm it}$ iterations, which gives $\Delta t = 0.13$ for $K=10$ in reduced units.

\begin{figure}[!ht]
  \centering
  \includegraphics{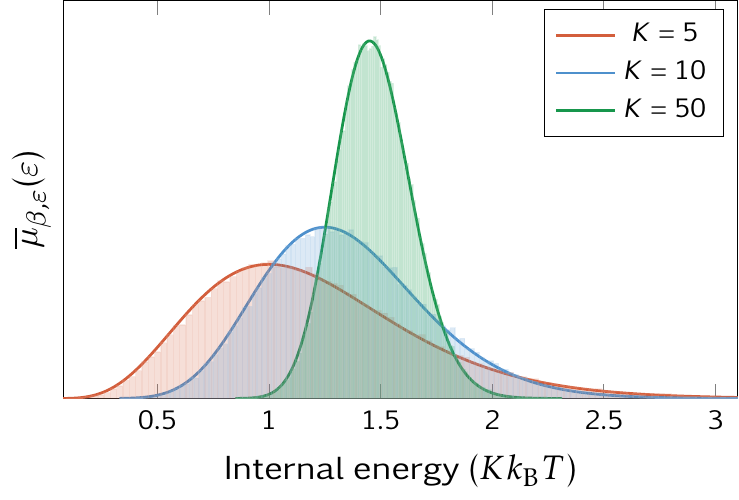}
  \caption{Internal energy distribution for the ideal gas rescaled by the size $K$ of the particles. The simulation results (histograms) are compared to analytic distribution (solid line). }
  \label{fig:pg-sdpd-eint}
\end{figure}

Under the invariant measure~\eqref{eq:sdpd-minveq}, the momenta are distributed according to a normal distribution with mean $0$ and variance $\displaystyle K\frac{m_0}{\beta}$ (so that the variance of the velocities scales as $\frac1K$) , which is well recovered in our simulations.
Moreover, thanks to the analytic form of the equation of state~\eqref{eq:pg-eos}, we can determine the theoretical distribution of the internal energies.
By integrating out the positions and momenta in the invariant measure~\eqref{eq:sdpd-minveq}, the marginal law for the internal energy is given by
\[
\overline{\mu}_{\beta,\varepsilon}(\rd \varepsilon) = \frac{ \beta^{ \frac{C_K}{k_{\rm B}}} }{ \Gamma\left(\frac{C_K}{k_{\rm B}}\right) }\varepsilon^{\frac{C_K}{k_{\rm B}}-1}\exp\left(-\beta\varepsilon\right)\,\rd\varepsilon,
\]
where $C_K = \frac32(K-1)k_{\rm B}$ is the heat capacity in the equation of state~\eqref{eq:pg-eos} and $\Gamma$ is the Gamma function.
We check that we recover this distribution in our simulations for $K=5$, $K=10$ and $K=50$ in Figure~\ref{fig:pg-sdpd-eint}.

We now study the evolution of the average pressure and temperature with respect to the particle size for the Lucy kernel~\eqref{eq:sdpd-lucy-w} and the cubic kernel~\eqref{eq:sdpd-cubic-w}.
The pressure is estimated according to~\eqref{eq:estim-pressure}.
There are two temperature estimators: the kinetic temperature~\eqref{eq:estim-tkin} and the internal temperature~\eqref{eq:estim-tint}.
We compare the simulation results to the values given by the equation of state~\eqref{eq:pg-eos}.
Small biases (around $0.1 \%$) are observed for small particle sizes but, as predicted by equations~\eqref{eq:estim-tkin} and~\eqref{eq:estim-tint}, the kinetic and internal temperatures are in excellent agreement with the equation of state as soon as $K \geq 100$.
There is no theoretical result predicting a perfect agreement of the pressure obtained by SDPD with the equation of state.
However, the results obtained from SDPD simulations with both the Lucy kernel and the cubic kernel match the pressure from the equation of state with a maximum of $5\%$ difference for sizes varying from $K=5$ to $K=25000$.
The thermodynamic limit for pressure is reached for sizes $K \geq 1000$. 
Let us however mention that we observe some metastability issues at large masses for the Lucy kernel due to particle clumping.
This leads us to prefer the cubic kernel in the following computations.

\captionsetup[subfigure]{position=top,captionskip=-10pt,margin=28pt}
\begin{figure}[!ht]
  \centering
  \subfloat[]{\includegraphics{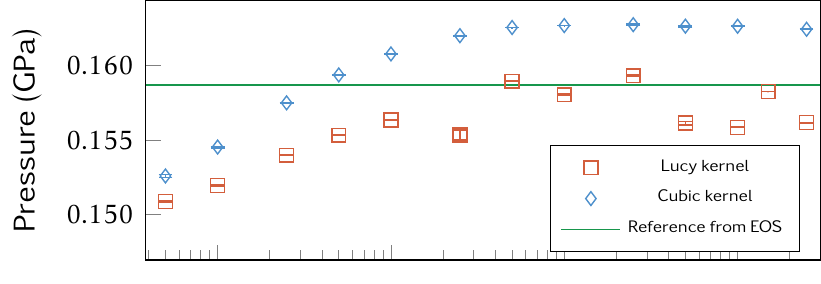}\label{fig:pg-consistency-pressure}}\\[-5pt]
  \subfloat[]{\includegraphics{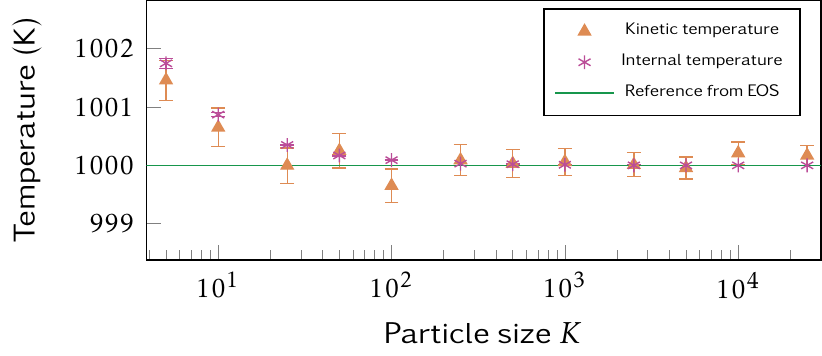}\label{fig:pg-consistency-temperature}}
  \caption{Numerical estimation of~\protect\subref{fig:pg-consistency-pressure} the equilibrium pressure, and~\protect\subref{fig:pg-consistency-temperature} the kinetic and internal temperatures as a function of the size $K$ of the SDPD particles (displayed with a logarithmic scale) for the ideal gas equation of state. Error bars are computed by integrating in time the autocorrelation as discussed in~\cite{lelievre_2016}.}
  \label{fig:pg-consistency}
\end{figure}

\subsection{Lennard-Jones fluid}
\label{sec:lj}
We discuss in this section the size consistency of properties estimated for a more realistic fluid of Lennard-Jones type. 
In a fully atomistic model, particles interactions can be modeled by a pairwise potential of Lennard-Jones type:
\[
\mathcal{U}_{\rm LJ}(r) = 4 \varepsilon_{\rm LJ} \left[\left(\frac{\sigma_{\rm LJ}}{r}\right)^{12} - \left(\frac{\sigma_{\rm LJ}}{r}\right)^6\right].
\]
We use the standard parameters for Argon ($\sigma_{\rm LJ} = 3.405\times10^{-10}$~m, $\varepsilon_{\rm LJ} = 1.657\times10^{-21}$~J, $m_0 = 6.64\times{10}^{-25}$~kg).
For SDPD, we use the equation of state for Lennard-Jones fluids presented in~\cite{johnson_1993}.
It is based on microscopic simulations carried out with MD in the NVT ensemble.
The Helmholtz free energy $\mathscr{F}(\rho,T)$ is fitted as a function of density and temperature on a modified form of the Benedict-Webb-Rubin equation of state~\cite{jacobsen_1973}, with 33 parameters.
The useful quantities for SDPD, like the internal energy $\mathcal{E}(\rho,T)$, the entropy $\mathcal{S}(\rho,T)$, the pressure $\mathcal{P}(\rho,T)$ and the heat capacity $\mathcal{C}(\rho,T)$, are then deduced from the free energy as
\[
  \begin{aligned}
    \mathcal{E}(\rho,T) &= -T^2 \partial_T\left(\frac{\mathscr{F}(\rho,T)}{T}\right),\\
    \mathcal{S}(\rho,T) &= \frac{\mathcal{E}(\rho,T)-\mathscr{F}(\rho,T)}{T},\\
    \mathcal{P}(\rho,T) &= \rho^2\partial_{\rho}\mathscr{F}(\rho,T),\\
    \mathcal{C}(\rho,T) &= \partial_T\mathcal{E}(\rho,T).
  \end{aligned}
\]
Since we use the internal energy as our primary variable, we perform a Newton inversion algorithm to find the temperature corresponding to a given internal energy $\varepsilon_i$ and density $\rho_i$ whenever we need to compute the associated pressure or temperature.
Denoting the temperature at iteration $k$ by $T_i^k$, we initialize the algorithm with $T_i^0 = T_{\beta}$, with $T_{\beta}$ the thermodynamic temperature, and iterate until the relative residual
\[
  \frac{\mathcal{E}(\rho_i,T^k_i) - \varepsilon_i}{\varepsilon_i}
\]
decreases below a threshold $\kappa_{\rm tol} = 10^{-9}$.
Usually only a few iterations are required for convergence.

The functional form of the equation of state~\cite{johnson_1993} diverges for small temperatures.
Since a few particles can, at few occasions, reach this small temperature domain due to fluctuations, we continuously extend the equation of state for $T<T_{\rm low}$ by choosing the heat capacity to be independent of temperature, \emph{i.e.}:
\[
  \mathcal{C}(\rho,T) = \mathcal{C}(\rho,T_{\rm low}), \text{ if } T < T_{\rm low}.
\]
We then write the energy and entropy in the regime $T < T_{\rm low}$ as
\[
  \begin{aligned}
    \mathcal{E}(\rho,T) &= \mathcal{C}(\rho,T_{\rm low})(T-T_{\rm low}) + \mathcal{E}(\rho,T_{\rm low})\\
    \mathcal{S}(\rho,T) &= \mathcal{C}(\rho,T_{\rm low})\log\left(\frac{T}{T_{\rm low}}\right) + \mathcal{S}(\rho,T_{\rm low}).
  \end{aligned}
\]
This enhances the stability of our simulations where we use $T_{\rm low} = 100$~K. 

We run simulations for systems of $N=1000$ particles initialized at temperature $T=1000$~K and density $\rho=1150$~kg.m$^{-3}$ as in Section~\ref{sec:vnum}.
We use the cubic kernel and the reduced units defined in Section~\ref{sec:scaling-sdpd}.
The number of iterations is fixed to $N_{\rm it}=5\times10^5$.
The timestep is chosen for each $K$ such that the drift in energy is less than $0.5 \%$ after $N_{\rm it}$ iterations, which gives $\Delta t = 0.03$ for $K=10$ in reduced units.

\begin{figure}[!ht]
  \centering
  \subfloat[]{\includegraphics{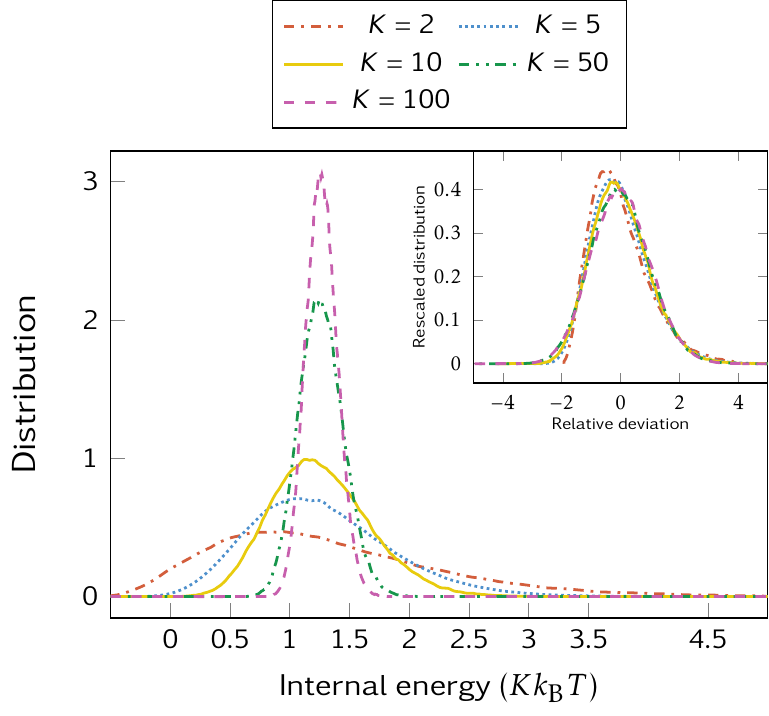}\label{fig:ljc_diste}}\\[-5pt]
  \captionsetup[subfigure]{position=top,captionskip=-22pt,margin=21pt}
  \subfloat[]{\includegraphics{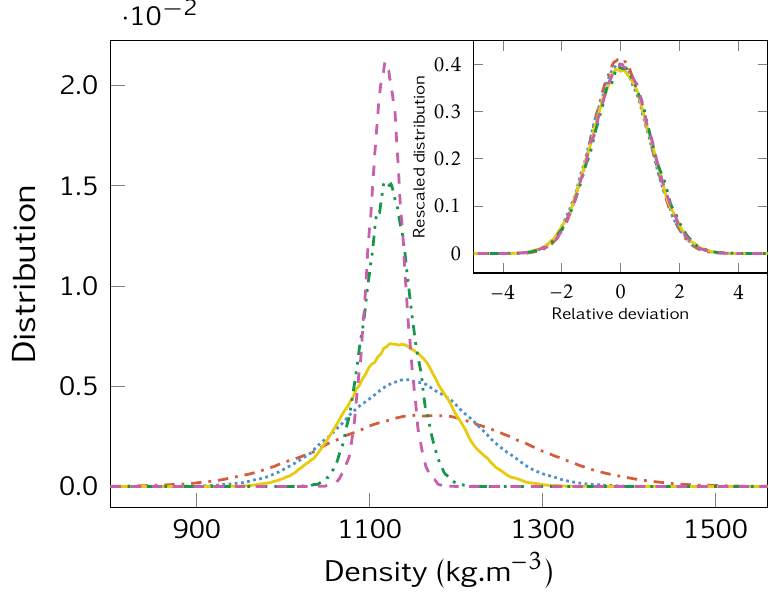}\label{fig:ljc_distd}}\\
  \captionsetup[subfigure]{position=top,captionskip=-10pt,margin=21pt}
  \subfloat[]{\includegraphics{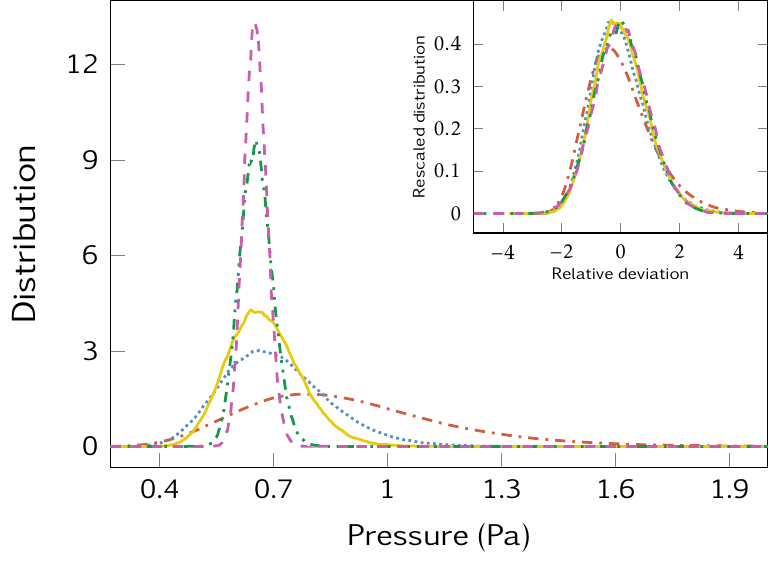}\label{fig:ljc_distp}}
  \caption{Distributions, for different SDPD masses $K$, of~\protect\subref{fig:ljc_diste} the normalized internal energy $\displaystyle \frac{\varepsilon_i}{K}$,~\protect\subref{fig:ljc_distd} the density $\rho_i$ and~\protect\subref{fig:ljc_distp} the internal pressure $P_i$. The rescaled distributions are displayed as insets in these figures.}
  \label{fig:ljc-dist}
\end{figure}

We plot the distributions of normalized internal energies $\displaystyle \frac{\varepsilon_i}{K}$, densities $\rho_i$ and pressures $P_i$ for several masses (see Figure~\ref{fig:ljc-dist}).
For each distribution, we find the appropriate scaling of their expected values $\mathfrak{m}_K$ and variance $\mathfrak{S}_K^2$ with respect to the particle size $K$ by fitting $\mathfrak{m}_K$ as a second-order polynomial in $\displaystyle \frac1K$ and $\mathfrak{S}_K^2$ as a power law in $K$ (see Table~\ref{tab:rescaling-dist}).
\begin{table}
  \centering
  {\renewcommand{\arraystretch}{1.8}
  \begin{tabular}{|c|c|c|}
    \hline
    & Expected value $\mathfrak{m}_K$ & Variance $\mathfrak{S}_K$ \\
    \hline
    Energy & $\displaystyle 1.266 - \frac{0.088}{K} + \frac{0.394}{K^2}$ & $\displaystyle 1.81K^{-1.008}$ \\[.5em]
    Density & $\displaystyle 1116 + \frac{187}{K} -\frac{167}{K^2}$ & $\displaystyle 2.5\times10^{4}K^{-0.924}$ \\[.5em]
    Pressure & $\displaystyle 0.649 + \frac{0.178}{K} + \frac{0.612}{K^2}$ & $\displaystyle 0.114K^{-1.004}$ \\[.5em]
    \hline
  \end{tabular}}
  \caption{Scaling of the expected value and variance for internal energy, density and pressure with respect to the particle size $K$.}
  \label{tab:rescaling-dist}
\end{table}
\captionsetup[subfigure]{position=bottom,captionskip=-163pt,margin=21pt}

The distributions are then rescaled as 
\[
\widetilde{f}_K(x) = \mathfrak{S}_K f_K(\mathfrak{m}_K+\mathfrak{S}_K x),
\]
where $f_K$ is the distribution function for some quantity.
The rescaled distributions $\widetilde{f}_K$ are represented as an inset in Figure~\ref{fig:ljc-dist}.
We notice that, for $K>5$, the rescaled distributions $\widetilde{f}_K$ collapse to a single distribution independent of the particle size $K$.
As $K$ increases, the mesoparticles stand for a larger collection of molecules, so that some effective averaging process takes place.
Standard results from probability theory suggest that the variances $\mathfrak{S}_K$ scale as $\displaystyle\frac1{K}$ and that the distribution should tend to a normal distribution.
Our numerical results are in excellent agreement with this prediction, the distributions becoming more symmetric for larger sizes $K$ with a variance inversely proportional to $K$.

\begin{figure}[!ht]
  \centering
  \subfloat[]{\includegraphics{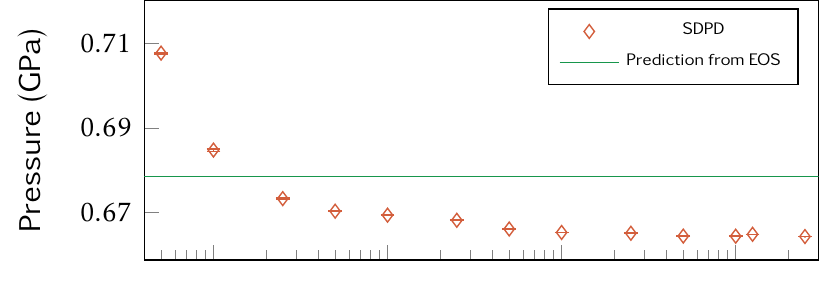}\label{fig:lj-consistency-pressure}}\\[-5pt]
  \subfloat[]{\includegraphics{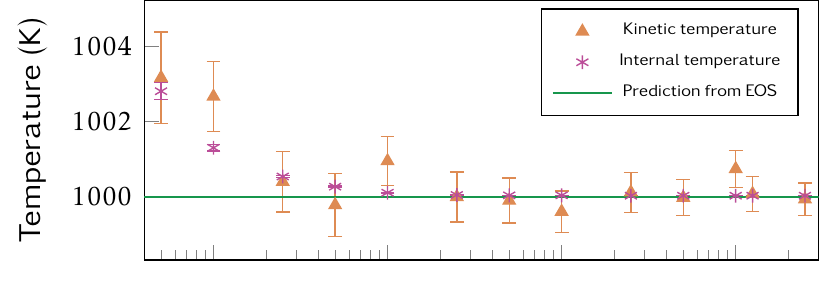}\label{fig:lj-consistency-temperature}}\\[-5pt]
  \subfloat[]{\includegraphics{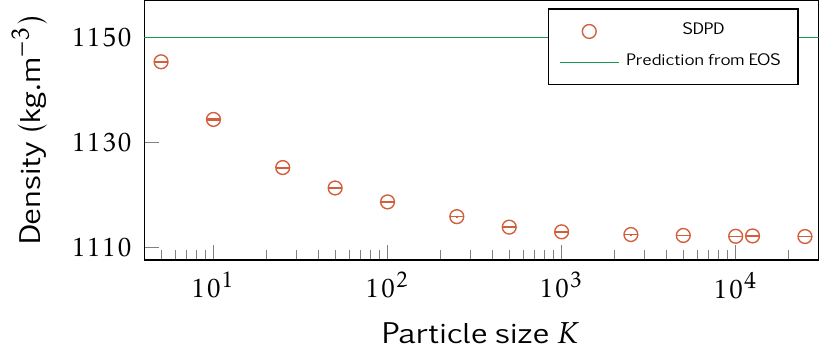}\label{fig:lj-consistency-density}}
  \caption{Numerical estimations of~\protect\subref{fig:lj-consistency-pressure} the equilibrium pressure,~\protect\subref{fig:lj-consistency-temperature} the kinetic and internal temperatures and~\protect\subref{fig:lj-consistency-density} the density as a function of the size $K$ of the SDPD particles (displayed with a logarithmic scale) for the Lennard-Jones equation of state.}
  \label{fig:lj-consistency}
\end{figure}

We check the consistency of the SDPD simulations with the reference equation of state by plotting the average pressure, internal and kinetic temperatures with respect to the mass of the fluid particles in Figure~\ref{fig:lj-consistency}.
\captionsetup[subfigure]{position=top,captionskip=-10pt,margin=28pt}
For large masses, the thermal fluctuations included in the SDPD equations are no longer significant and the estimated pressure converges to some limiting pressure which is $3\%$ lower than the pressure predicted by the equation of state.
However, we also note a discrepancy in the estimation of the density as we observe an average mean density of $1112$~kg.m$^{-3}$ for $K>1000$ instead of the expected value $1150$~kg.m$^{-3}$.
As the size of the particles decreases, the agreement between the pressure obtained from SDPD simulations and the pressure predicted by the equation of state remains within $5\%$ of difference even for $K=5$ or $K=10$, \emph{i.e.} for masses of the order of only a few multiples of $m_0$.

\subsection{Shock waves}
\label{sec:shocks}

We turn in this section to the study of the consistency of SDPD in nonequilibrium situations such as shock waves.
We consider a system of $N=54272$ particles initialized on a simple cubic lattice $16\times16\times212$.
We use periodic boundary condition in the $x$- and $y$-directions whereas two walls are located at each end of the system in the $z$-direction.
Each wall is composed of 3 layers of ``virtual'' SDPD particles arranged in a cubic lattice, using the same ideas as Bian \emph{et al.}~\cite{bian_2012}.
The positions and momenta of the virtual particles are not updated with the dynamics but are kept fixed within the walls.
These virtual particles enable us to evaluate the density of the SDPD particles in the neighborhood of the walls, as well as the conservative forces~\eqref{eq:sdpd-nrj-cons} acting on the actual particles.
In order to ensure that walls are not permeable, these particles induce a repulsive force deriving from a truncated Lennard-Jones potential:
\[
\mathcal{U}_{\rm rLJ}(r) = 4 \varepsilon_{\rm rLJ} \left[\left(\frac{\sigma_{\rm rLJ}}{r}\right)^{12} - \left(\frac{\sigma_{\rm rLJ}}{r}\right)^6 + \frac14\right] \mathds{1}_{r \leq 2^{\frac16}\sigma_{\rm rLJ}}.
\]
We set $\varepsilon_{\rm rLJ} = 1$ and $\sigma_{\rm rLJ}=1$ in the reduced units~\eqref{eq:reduced-units}.

We use the Lennard-Jones equation of state~\cite{johnson_1993} in these simulations.
The system is initialized at a temperature $T_0=500$~K and a density $\rho_0=1150$~kg.m$^{-3}$.
To produce a sustained shock wave in the system, the bottom wall is set in motion at a velocity $v_P = 500$~m.s$^{-1}$ in the $z$-direction and continues moving at velocity $v_P$ throughout the simulation.
We run the simulation with particle sizes varying from $K=10$ to $K=10000$.

In order to avoid the effects due to the presence of the walls, we only consider the information arising from particles located at a distance larger than $10\sigma$ from a wall.
Making use of the stationarity of shock waves in the reference frame of the shock front, it is possible to average profiles over time.
We split the simulation box into a number of slices $n_{\rm sl} = 100$ regularly distributed along the $z$-axis and compute average quantities in the slices.
We determine the position of the shock front at every step as the point where the mean particles velocity along the $z$-axis is the closest to the velocity  $\frac{v_P}{2}$.
The various profiles are then averaged by setting the position of the shock front as the reference frame ($z=0$).

Table~\ref{tab:shock-500-ljv} summarizes the main physical properties estimated with the simulations: the velocity of the shock front $v_S$ along with the thermodynamic properties in the shocked state (the density $\rho_S$, the pressure $P_S$ and the internal temperature $T_{{\rm int},S}$).
They are compared to the corresponding values obtained via direct MD simulation and to the values predicted by the Rankine-Hugoniot relations applied to the equation of state~\cite{johnson_1993}.
Assuming that the evolution can be described in an effective manner by a one-dimensional Euler system (in particular, viscosity effects can be neglected), the Rankine-Hugoniot conditions allow to predict the thermodynamic properties in the shocked state, knowing the initial thermodynamic state, the velocity of the shock wave and the velocity of the particles in the shocked region.
These conditions are obtained from the conservation laws for mass, momentum and energy.
The density $\rho_S$, pressure $P_S$ and internal energy per unit mass $u_S$ in the shocked state are respectively predicted to be
\[
  \begin{aligned}
    \rho_S &= \rho_0\frac{v_S}{v_S-v_P},\\
    P_S &= P_0 + \rho_0 v_Sv_P,\\
    u_S-\frac12v_S^2+\frac{P_S}{\rho_S} &=u_0+\frac12v_0^2+\frac{P_0}{\rho_0},
  \end{aligned}
\]
where $\rho_0$, $P_0$ and $u_0$ are the density, pressure and internal energy per unit mass in the initial unshocked region.
We find that SDPD gives similar results for all the resolutions which are considered.
These results are consistent with the predictions obtained from the Rankine-Hugoniot relations.
They also agree with MD within a $5\%$-difference margin for pressure, density and temperature.
While no particular bias is observed for temperature, the density observed in SDPD is slightly higher and the pressure slightly lower than the MD results.
The shock velocity is also a bit underestimated and seems to decay for larger particles.
Since we consider a bulk material, the viscosity has no effect on the average properties in the shocked state.
As such, we obtain quasi identical results when $\eta=10^{-4}$~Pa.s (as presented in Table~\ref{tab:shock-500-ljv}) or $\eta = 2\times10^{-3}$~Pa.s

\begin{table}[htb]
  \centering
  \begin{tabular}{|c|c|c|c|c|}
    \hline
    $K$ & $v_S$ (m.s$^{-1}$)& $\rho_{S}$ (kg.m$^{-3}$) & $P_{S}$ (GPa) & $T_{{\rm int},S}$ (K)\\
    \hline
    MD & 1961 & 1508 & 1.45 & 939\\
    EoS & 1975 & 1540 & 1.47 & 969\\
    10 & 1846 & 1547 & 1.37 & 911\\
    100 & 1900 & 1546 & 1.39 & 946\\
    500 & 1897 & 1547 & 1.39 & 946\\
    1000 & 1886 & 1545 & 1.37 & 938\\
    5000 & 1870 & 1552 & 1.39 & 935\\
    10000 & 1864 & 1551 & 1.38 & 929\\
    \hline
  \end{tabular}
  \caption{\label{tab:shock-500-ljv}Average observables in the shocked state. The wall velocity is fixed to $v_{P}=500$~m.s$^{-1}$ and the viscosity parameter for SDPD is set to $\eta=10^{-4}$~Pa.s.}
\end{table}

Following~\cite{holian_1993,holian_2010_b}, we compute the one-dimensional Navier-Stokes shock wave profile.
In this 1D stationary setting, the Navier-Stokes conservation equations~\eqref{eq:navier-stokes} simplify to simple differential equations:
\begin{equation}
  \label{eq:1d-stationary-ns}
  \begin{aligned}
    \rho(z)v(z) &= \rho_0v_0,\\
    P(z)-\eta v'(z)+\rho_0v_0v(z) &=P_0+\rho_0v_0^2,\\
    u(z)-\frac12v(z)^2+\frac{P_0+\rho_0v_0^2}{\rho(z)}&=u_0+\frac12v_0^2+\frac{P_0}{\rho_0},
  \end{aligned}
\end{equation}
where $z\in\mathbb{R}$ is the distance to the shock front (located at $z=0$).
The unknowns are the density $\rho$, the velocity $v$, and the internal energy per unit mass $u$, while the pressure $P$ is given by the equation of state.
We choose the unshocked material to be at $z>0$  and the shocked fluid at $z<0$.
This determines the velocity in the unshocked state to be $v_0 = -v_S$ and the velocity in the shocked state $v_1 = v_P-v_S$.
Equations~\eqref{eq:1d-stationary-ns} are integrated with a finite differences scheme on a domain $[-L,L]$ and with initial conditions $v(0)=\frac12(v_0+v_S)$.
The density $\rho(0)$ and energy $u(0)$ at the origin are then determined thanks to equations~\eqref{eq:1d-stationary-ns}.
We choose here $L=10^{-8}$~m and a mesh spacing $\Delta x = 10^{-12}$~m.

We also present the results for a MD simulation of the same setting.
In MD simulations, the walls are modeled as infinitely massive particles interacting with the other particles through a Lennard-Jones potential.
We use a Velocity-Verlet scheme~\cite{verlet_1967} and a timestep $\Delta t = 10^{-15}$~s for the integration of the Hamiltonian dynamics.

\begin{figure}[!ht]
  \centering
  \subfloat[]{\includegraphics{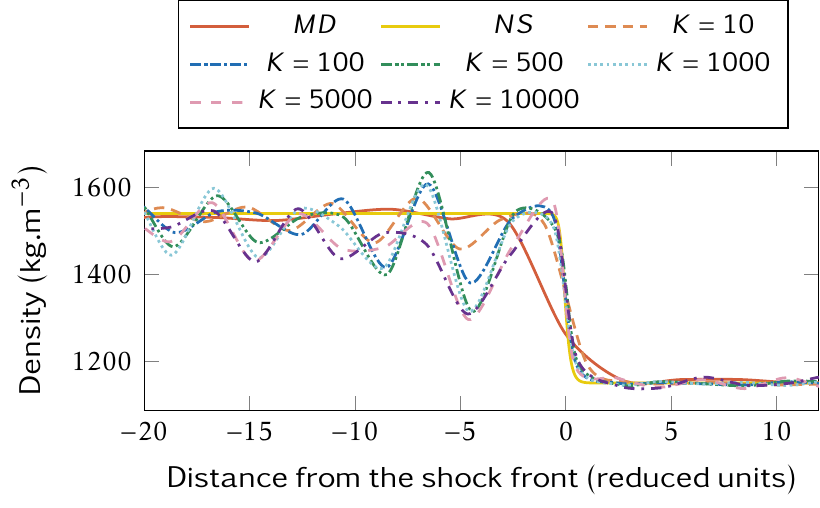}\label{fig:shock-profiles-reduced}}\\[0pt]
  \captionsetup[subfigure]{position=top,captionskip=-10pt,margin=28pt}
  \subfloat[]{\includegraphics{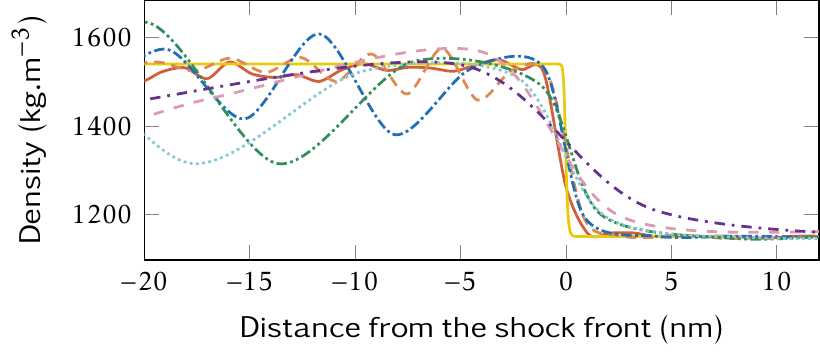}\label{fig:shock-profiles-nm}}
  \caption{Density profiles in the shock reference frame for $K=10$ to $K=10000$ compared to MD and Navier-Stokes (NS). The reduced units are defined by equation~(\ref{eq:reduced-units}). The viscosity is set to $\eta=10^{-4}$~Pa.s and the wall velocity to $v_P=500$~m.s$^{-1}$.}
  \label{fig:shock-profiles}
\end{figure}

The mean profiles for density are given in Figure~\ref{fig:shock-profiles} in the corresponding reduced units and in physical units.
The reduced length unit for MD and Navier-Stokes is the same as the SDPD reduced length unit for $K=1$ to allow for a comparison with the other profiles.
\captionsetup[subfigure]{position=bottom,captionskip=-103pt,margin=28pt}
Figure~\ref{fig:shock-profiles} is obtained by using a viscosity $\eta=10^{-4}$~Pa.s, which is of the same order as the viscosity of the Lennard-Jones fluid, for the SDPD simulations and the Navier-Stokes solution, which allows for a comparison between these methods.
While the profile derived from~\eqref{eq:1d-stationary-ns} is sharper than the MD profile, we observe that we can recover reasonably well the profile from MD for small SDPD particle sizes.
When the size of the SDPD particles increases, the shock front widens and no longer agrees with the MD profile.
Since the width of the shock front in SDPD seems constant in reduced units for any of the tested resolutions, the main factor governing the shock width in physical units in this situation appears to be the resolution chosen for SDPD.

\begin{figure}[!ht]
  \centering
  \subfloat[]{\includegraphics{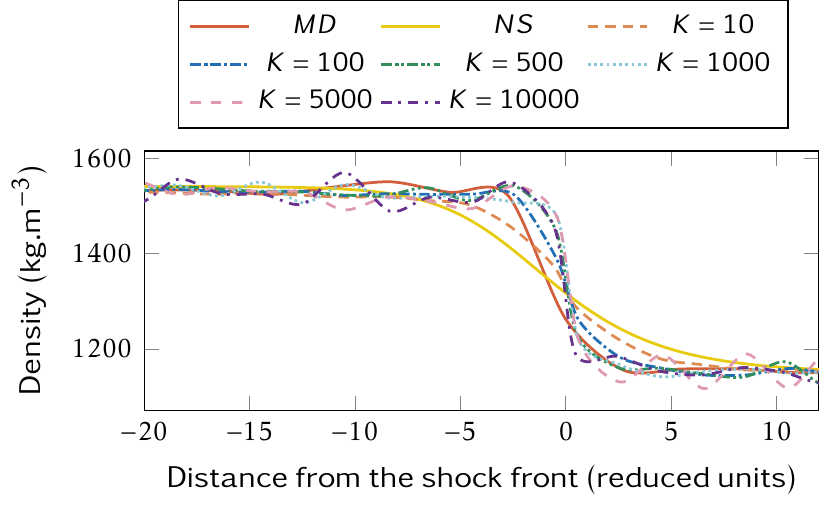}\label{fig:shock-profiles-viscous-reduced}}\\[0pt]
  \captionsetup[subfigure]{position=top,captionskip=-10pt,margin=28pt}
  \subfloat[]{\includegraphics{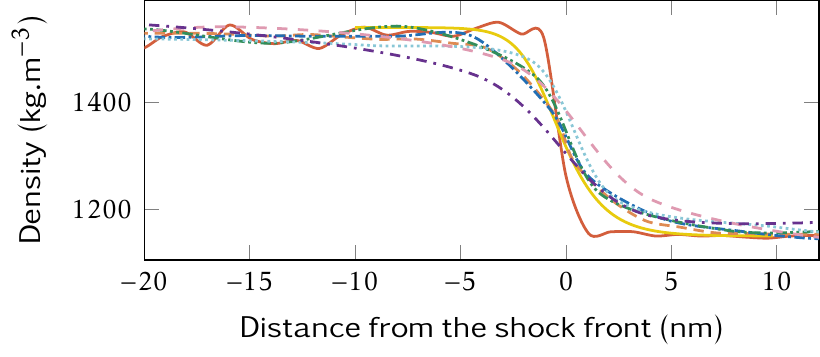}\label{fig:shock-profiles-viscous-nm}}
  \caption{Density profiles in the shock reference frame for $K=10$ to $K=10000$ compared to MD and Navier-Stokes (NS). The reduced units are defined by equation~(\ref{eq:reduced-units}). The viscosity is set to $\eta=2\times10^{-3}$~Pa.s and the wall velocity to $v_P=500$~m.s$^{-1}$.}
  \label{fig:shock-artificial-visco}
\end{figure}

The profiles computed with SDPD display strong oscillations in the shocked state due to the small value of the viscosity.
Similar issues are encountered in SPH, where an artificial viscosity is introduced to alleviate the oscillations~\cite{monaghan_1983}.
Figure~\ref{fig:shock-artificial-visco} presents the profiles computed with a larger viscosity $\eta=2\times10^{-3}$~Pa.s, which is comparable to water.
The oscillations are effectively dampened but the shock front is now quite wide compared to MD and agrees with the Navier-Stokes results.
Its width in physical units no longer depends on the particles size for the range of resolution we study.
In this situation, it appears that the dominating effect is the viscosity and any of the tested particles sizes is able to accurately resolve the shock front for moderately viscous fluids.
We anticipate similar results for higher viscosity fluids.
\captionsetup[subfigure]{position=bottom,captionskip=-103pt,margin=28pt}

\section{Conclusion}
\label{sec:conclusion}
We presented in this work a reformulation of the SDPD equations in terms of internal energies rather than internal entropies.
This leads to a set of stochastic differential equations with a structure very similar to that of the DPDE equations, which opens the way for a concurrent coupling of the two methods.
It also enables us to integrate the SDPD equations with energy preserving numerical schemes originally developed for DPDE.

In order for SDPD to reproduce the behavior of a microscopic system, we chose to use an equation of state fitted on equilibrium MD simulations for a Lennard-Jones fluid~\cite{johnson_1993}.
Using this equation of state, we studied the influence of the resolution level in SDPD on the fluid properties.
We showed that the equilibrium thermodynamic properties can be retrieved by SDPD for a wide range of particle sizes, even down to the scale of microscopic particles.
Moreover, the distribution functions for the particles density, internal temperature and pressure behave as Gaussian distributions with a variance scaling as $\displaystyle\frac1{K}$ when the size of the particle increase.
This is consistent with the viewpoint that mesoparticles represent the average behavior of $K$ underlying microscopic particles.

We finally tested the method in a non equilibrium situation, namely a shock wave, and compared the resulting profiles for MD and SDPD at different resolutions.
We found that the thermodynamic properties in the shocked state agree quite well with the predictions of MD with errors of the same order as for equilibrium properties.
As far as nonequilibrium properties are concerned, two regimes can be distinguished: for low viscosity fluids like argon, the width of the shock front in physical units is governed by the size of the SDPD particles, and the agreement with MD profiles is only recovered for very small sizes ($K<10$).
Moreover, spurious oscillations appear behind the shock front.
For higher viscosity fluids (\emph{i.e.} water or above), the width of the shock front in physical units become independent of the resolution of SDPD simulations and is controlled by the viscosity.
Oscillations are also damped out.

The consistency of the results at the various levels of coarse-graining, as well as the agreement with the original, atomistic system, allows us to envision a concurrent coupling of SDPD at different resolutions.

\section*{Acknowledgments}
We thank Pep Espa\~nol for fruitful discussions and John Brennan for bringing~\cite{johnson_1993} to our attention.
The work of G.S. was funded by the Agence Nationale de la Recherche, under grant ANR-14-CE23-0012 (COSMOS) and by the European Research Council under the European Union's Seventh Framework Programme (FP/2007-2013) / ERC Grant Agreement number 614492.

\appendix
\section{Energy conservation for the reformulated SDPD}
\label{sec:app-cons}
We check in this appendix that the dynamics~\eqref{eq:sdpd-energy} preserves the energy $\displaystyle E(\vect{q},\vect{p},\varepsilon) = \sum\limits_{i=1}^N \frac{\vect{p}_i^2}{2m} + \sum\limits_{i=1}^N \varepsilon_i$ by showing that each elementary dynamics~\eqref{eq:sdpd-nrj-cons} and~\eqref{eq:sdpd-simple-fluct} independently preserve the total energy $E(\vect{q},\vect{p},\varepsilon)$.
The variation of the energy along the trajectory is given by
\begin{equation}
  \label{eq:nrj-variation}
  \rd E(\vect{q},\vect{p},\varepsilon) = \sum\limits_{i=1}^N\frac1{2m} \rd\left(\vect{p}_i^2\right) + \rd \varepsilon_i.
\end{equation}
We next evaluate this variation for the sub-dynamics under consideration.
\subsection{Conservation of the energy by the conservative part of the dynamics}
\label{sec:cons-energy-cons}
We first study the conservative dynamics~\eqref{eq:sdpd-nrj-cons}.
Since it is a deterministic dynamics, the energy variation simply reads
\[
  \rd E(\vect{q},\vect{p},\varepsilon) = \sum\limits_{i=1}^N\frac1{m} \vect{p}_i\cdot \rd\vect{p}_i + \rd \varepsilon_i.
\]
In view of the equations of motion~\eqref{eq:sdpd-nrj-cons}, we get
\[
  \begin{aligned}
    \rd E(\vect{q},\vect{p},\varepsilon) =& \sum\limits_{i=1}^N\frac1{m} \vect{p}_i\cdot \left[ \sum_{j\neq i}m^2\left(\frac{P_i}{\rho_i^2}+\frac{P_j}{\rho_j^2}\right)F_{ij}\vect{r}_{ij}\,\rd t \right]\\
    &- \sum\limits_{i=1}^N\sum_{j\neq i}m^2\frac{P_i}{\rho_i^2}F_{ij}\vect{r}_{ij}\cdot\vect{v}_{ij}\,\rd t.\\
  \end{aligned}
\]
A final reorganization of the second term makes it clear that the right hand side vanishes and proves the conservation of the energy by the elementary dynamics~\eqref{eq:sdpd-nrj-cons}.

\subsection{Conservation of the energy by the fluctuation and dissipation part of the dynamics}
\label{sec:cons-energy-fluct}
We now focus on the elementary pairwise fluctuation and dissipation dynamics~\eqref{eq:sdpd-simple-fluct} for a given pair $(i,j)$.
This dynamics is stochastic.
Itô calculus yields
\[
  \begin{aligned}
    \rd\left[\frac{\vect{p}_i^2}{2m} + \frac{\vect{p}_j^2}{2m}\right] &= \frac{\vect{p}_i}m\cdot \rd\vect{p}_i + \frac{\vect{p}_j}m\cdot \rd\vect{p}_j + \frac{{\rm Tr}\left(\mtx{\Sigma}_{ij}\mtx{\Sigma}_{ij}^T\right)}{m}\rd t\\
    &= \vect{v}_{ij}\cdot \rd\vect{p}_i + \frac1m{\rm Tr}\left(\mtx{\Sigma}_{ij}\mtx{\Sigma}_{ij}^T\right)\rd t.
  \end{aligned}
\]
In view of the equations of motion~\eqref{eq:sdpd-simple-fluct}, the variation of the internal energies can be rewritten as
\[
  \rd\left(\varepsilon_i + \varepsilon_j\right) = -\vect{v}_{ij}\cdot \rd\vect{p}_i - \frac1{m}{\rm Tr}\left(\mtx{\Sigma}_{ij}\mtx{\Sigma}_{ij}^T\right)\rd t.
\]
This proves that $\rd E(\vect{q},\vect{p},\varepsilon) = 0$.
The energy is thus conserved by the pairwise dynamics~\eqref{eq:sdpd-simple-fluct}.

Since all the elementary dynamics preserve the energy, the global dynamics~\eqref{eq:sdpd-energy} obtained by the superposition of~\eqref{eq:sdpd-nrj-cons} and~\eqref{eq:sdpd-simple-fluct} also preserves the energy $E(\vect{q},\vect{p},\varepsilon)$.

\section{Invariant measure for the reformulated SDPD}
\label{sec:app-minv}
We check that measures $\mu$ of the form~\eqref{eq:sdpd-energy-minv} are left invariant by the SDPD dynamics~\eqref{eq:sdpd-energy}.
We proceed by showing that $\mu$ is invariant by the conservative dynamics~\eqref{eq:sdpd-nrj-cons} in Section~\ref{sec:minv-cons} and by the elementary pairwise fluctuation/dissipation dynamics~\eqref{eq:sdpd-simple-fluct} in Section~\ref{sec:minv-fd}.
We however need some preliminary material to this end, which we provide in Section~\ref{sec:minv-and-gen} and~\ref{sec:minv-deriv}.

\subsection{The Fokker-Plank equation}
\label{sec:minv-and-gen}

In this section, we present some standard tools which we use in the following sections to study the SDPD stochastic differential equations and to prove the invariance of~\eqref{eq:sdpd-energy-minv}.
We consider stochastic dynamics of the form
\begin{equation}
  \label{eq:general-sde}
  \rd X_t = b(X_t)\,\rd t + \mathscr{S}(X_t)\rd\mathcal{W}_t,
\end{equation}
where the variable $X_t$ is of dimension $d$, the drift coefficient $b$ is a vector of dimension $d$, the fluctuation amplitude $\mathscr{S}$ a matrix of dimension $d\times n$ and $\mathcal{W}$ a standard Brownian motion of dimension $n$.
We can associate to the dynamics~\eqref{eq:general-sde} an operator $\mathcal{L}$, called the infinitesimal generator:
\[
  \mathcal{L} = b\cdot\grad_{X} + \frac12 \mathscr{S}\mathscr{S}^T:\grad_{X}^2,
\]
where the contraction operation for two matrices $A$ and $B$ of size $d\times d$ is defined as
\[
  A:B = \sum_{1\leq i,j\leq d} A_{ij}B_{ij}.
\]

We define the adjoint $\mathcal{A}^*$ of an operator $\mathcal{A}$ as the operator such that, for any $\phi$ and $\psi$, smooth and compactly supported test functions,
\[
\int_{\mathcal{E}}\left(\mathcal{A}\phi\right)\psi = \int_{\mathcal{E}} \phi\left(\mathcal{A}^*\psi\right).
\]
The adjoint $\mathcal{L}^*$ of the generator governs the evolution of the law $\psi(t)$ of the process $X_t$, solution of~\eqref{eq:general-sde} through the well-known Fokker-Plank equation as
\[
  \partial_t\psi = \mathcal{L}^*\psi.
\]
Thus, $\psi$ is a stationary solution of the Fokker-Plank equation if and only if
\[
  \mathcal{L}^*\psi = 0.
\]

In Sections~\ref{sec:minv-cons} and~\ref{sec:minv-fd}, we study the elementary sub-dynamics by writing the associated infinitesimal generators $\mathcal{L}_{\rm cons}$ and $\mathcal{L}_{{\rm fd},ij}$ along with their adjoints.
In order to prove the invariance of~\eqref{eq:sdpd-energy-minv}, we introduce $f_{\mu}$ the density of the measure $\mu$ (see~\eqref{eq:minv_density} below) and compute $\mathcal{L}_{\rm cons}^*f_{\mu}$ and $\mathcal{L}_{{\rm fd},ij}^*f_{\mu}$.
Since these operators are differential operators, we first need to evaluate the derivatives of the density function $f_{\mu}$, which we do in Section~\ref{sec:minv-deriv}.

\subsection{Evaluation of the derivatives of the density function}
\label{sec:minv-deriv}

We introduce a smooth function $g(E,\ptot)$ (with $E\in \mathbb{R}$ and $\ptot \in \mathbb{R}^3$) and the function
\[
  \begin{aligned}
    h(\rho,\varepsilon) &= \prod_{i=1}^N \frac1{\mathcal{T}(\varepsilon_i,\rho_i)}\exp\left(\frac{\mathcal{S}(\varepsilon_i,\rho_i)}{k_{\rm B}}\right)\\
    &= \prod_{i=1}^N \partial_{\varepsilon}\mathcal{S}(\varepsilon_i,\rho_i)\exp\left(\frac{\mathcal{S}(\varepsilon_i,\rho_i)}{k_{\rm B}}\right),\\
  \end{aligned}
\]
where we made use of the relations~\eqref{eq:eos-thermo}.
We introduce the following notation:
\[
  \begin{aligned}
    \mathfrak{h}(\vect{q},\vect{p},\varepsilon) &= h(\rho_1(\vect{q}),\dots,\rho_N(\vect{q}),\varepsilon),\\
    \mathfrak{g}(\vect{q},\vect{p},\varepsilon) &= g\left(E(\vect{q},\vect{p},\varepsilon),\sum\limits_{i=1}^N \vect{p}_i\right),
  \end{aligned}
\]
so that we can write the measure $\mu$ in~(\ref{eq:sdpd-energy-minv}) as
\[
  \mu(\rd\vect{q}\,\rd\vect{p}\,\rd \varepsilon) = f_{\mu}(\vect{q},\vect{p},\varepsilon)\,\rd\vect{q}\,\rd\vect{p}\,\rd \varepsilon,
\]
with
\begin{equation}
  \label{eq:minv_density}
  f_{\mu} (\vect{q},\vect{p},\varepsilon) = \mathfrak{g}(\vect{q},\vect{p},\varepsilon)\mathfrak{h}(\vect{q},\vect{p},\varepsilon).
\end{equation}
In order to compute the derivatives of $f_{\mu}$ with respect to the variables $\vect{q}_i$, $\vect{p}_i$ and $\varepsilon_i$, we first evaluate the derivatives of $\mathfrak{g}$ and $\mathfrak{h}$ with respect to these variables.
Since the total energy $\displaystyle E(\vect{q},\vect{p},\varepsilon) = \sum\limits_{i=1}^N \frac{\vect{p}_i^2}{2m} + \varepsilon_i$ does not depend on $\vect{q}$, the derivatives of $\mathfrak{g}$ read
\begin{equation}
  \label{eq:g-derivatives}
  \begin{aligned}
    \grad_{\vect{q}_i} \mathfrak{g}\left(E(\vect{q},\vect{p},\varepsilon),\sum\limits_{i=1}^N\vect{p}_i \right) &= 0,\\
    \grad_{\vect{p}_i} \mathfrak{g}\left(E(\vect{q},\vect{p},\varepsilon),\sum\limits_{i=1}^N\vect{p}_i \right) &= \frac1{m}\vect{p}_i\partial_E g + \grad_{\ptot} g,\\
    \partial_{\varepsilon_i} \mathfrak{g}\left(E(\vect{q},\vect{p},\varepsilon),\sum\limits_{i=1}^N\vect{p}_i \right) &= \partial_E g.
  \end{aligned}
\end{equation}
We note that $\mathfrak{h}$ does not depend on the momenta $\vect{p}_i$ and actually only depends on the positions $\vect{q}_i$ through the densities $\rho_i$.
We therefore first compute the derivatives of $h$ with respect to $\rho_i$ and $\varepsilon_i$:
\begin{equation}
  \label{eq:h-derivatives}
  \begin{aligned}
    \partial_{\rho_i} h(\rho,\varepsilon) =& \left[\partial_{\varepsilon_i} \mathcal{S}(\varepsilon_i,\rho_i)\partial_{\rho_i}\left(\exp\left(\frac{\mathcal{S}(\varepsilon_i,\rho_i)}{k_{\rm B}}\right)\right)\right.\\
    &\quad \left. +\, \partial_{\rho_i}\partial_{\varepsilon_i} \mathcal{S}(\varepsilon_i,\rho_i) \exp\left(\frac{\mathcal{S}(\varepsilon_i,\rho_i)}{k_{\rm B}}\right)\right]\\
    &\times \prod\limits_{j\neq i}  \partial_{\varepsilon}\mathcal{S}(\varepsilon_i,\rho_i)\exp\left(\frac{\mathcal{S}(\varepsilon_i,\rho_i)}{k_{\rm B}}\right)\\
    =& \left(\frac{\partial_{\rho_i}\mathcal{S}(\varepsilon_i,\rho_i)}{k_{\rm B}}+\frac{\partial_{\rho_i}\partial_{\varepsilon_i}\mathcal{S}(\varepsilon_i,\rho_i)}{\partial_{\varepsilon_i}\mathcal{S}(\varepsilon_i,\rho_i)}\right)h(\rho,\varepsilon),\\[.5em]
    \partial_{\varepsilon_i} h(\rho,\varepsilon) =& \left[\partial_{\varepsilon_i} \mathcal{S}(\varepsilon_i,\rho_i)\partial_{\varepsilon_i}\left(\exp\left(\frac{\mathcal{S}(\varepsilon_i,\rho_i)}{k_{\rm B}}\right)\right)\right.\\
    &\quad\left. +\, \partial_{\varepsilon_i}^2 S(\rho_i,\varepsilon_i) \exp\left(\frac{\mathcal{S}(\varepsilon_i,\rho_i)}{k_{\rm B}}\right)\right]\\
    &\times \prod\limits_{j\neq i}  \partial_{\varepsilon}\mathcal{S}(\varepsilon_i,\rho_i)\exp\left(\frac{\mathcal{S}(\varepsilon_i,\rho_i)}{k_{\rm B}}\right)\\
    =& \left(\frac{\partial_{\varepsilon_i}\mathcal{S}(\varepsilon_i,\rho_i)}{k_{\rm B}}+\frac{\partial_{\varepsilon_i}^2\mathcal{S}(\varepsilon_i,\rho_i)}{\partial_{\varepsilon_i}\mathcal{S}(\varepsilon_i,\rho_i)}\right)h(\rho,\varepsilon).
  \end{aligned}
\end{equation}
Equation~\eqref{eq:gradient-rho} allow us to express the derivatives of $\mathfrak{h}$ with respect to the positions in terms of the derivatives of $h$ with respect to the densities $\rho_i$ as
\begin{equation}
  \label{eq:hder-rho2q}
  \begin{aligned}
    \grad_{\vect{q}_i} \mathfrak{h}(\vect{q},\vect{p},\varepsilon) &= \left(\grad_{\vect{q}_i}\rho_i\right)\partial_{\rho_i}h +  \sum_{j\neq i} \left(\grad_{\vect{q}_i}\rho_j\right) \partial_{\rho_j}h\\
    &= -m \sum\limits_{j\neq i} (\partial_{\rho_i}h + \partial_{\rho_j}h) F_{ij}\vect{r}_{ij}.
  \end{aligned}
\end{equation}
Using equations~\eqref{eq:g-derivatives},~\eqref{eq:h-derivatives} and~\eqref{eq:hder-rho2q}, we are finally able to write the derivatives of $f_{\mu}$ with respect to the positions $\vect{q}_i$ as
\[
  \begin{aligned}
    \grad_{\vect{q}_{i}} f_{\mu} =&\, \mathfrak{g}\left(\vect{q},\vect{p},\varepsilon\right) \grad_{\vect{q}_i} \mathfrak{h}(\vect{q},\vect{p},\varepsilon)\\
    =& - m\mathfrak{g}\left(\vect{q},\vect{p},\varepsilon\right) \sum_{j \neq i} F_{ij}\vect{r}_{ij} (\partial_{\rho_j}h+\partial_{\rho_i}h);
  \end{aligned}
\]
the derivatives of $f_{\mu}$ with respect to the momenta $\vect{p}_i$ as
\[
  \grad_{\vect{p}_{i}} f_{\mu} = \mathfrak{h}(\vect{q},\vect{p},\varepsilon) \left[ \grad_{\ptot} g  + \frac{\vect{p}_i}{m} \partial_Eg \right],
\]
and the derivatives with respect to the energies $\varepsilon_i$ as
\[
  \partial_{\varepsilon_{i}} f_{\mu} = \mathfrak{g}\left(\vect{q},\vect{p},\varepsilon\right) \partial_{\varepsilon_i} h + \mathfrak{h}(\vect{q},\vect{p},\varepsilon) \partial_E g.
\]

\subsection{Invariance by the conservative part of the dynamics}
\label{sec:minv-cons}

The generator $\mathcal{L}_{\rm cons}$ associated with the dynamics~\eqref{eq:sdpd-nrj-cons} reads
\[
  \begin{aligned}
    \mathcal{L}_{\rm cons} = &\sum_{i=1}^N\frac{\vect{p_{i}}}{m}\cdot\grad_{\vect{q}_i} - m^2\sum_{i=1}^N\sum_{j \neq i}\frac{P_i}{\rho_i^2} F_{ij}\vect{r}_{ij}\cdot\vect{v}_{ij}\partial_{\varepsilon_i}\\
    &+ m^2\smashoperator[l]{\sum\limits_{1\leq i < j \leq N}}\left[\frac{P_i}{\rho_i^2}+\frac{P_j}{\rho_j^2}\right]F_{ij}\vect{r}_{ij}\cdot(\grad_{\vect{p}_i}-\grad_{\vect{p}_j}). 
  \end{aligned}
\]
In order to simplify the notation, we introduce
\[
  \mathfrak{F}_i = \sum\limits_{j\neq i} F_{ij}\vect{r}_{ij}\cdot\vect{v}_{ij}.
\]
The adjoint of the generator $\mathcal{L}_{\rm cons}$ is readily given by
\[
  \begin{aligned}
    \mathcal{L}_{\rm cons}^*\phi =& \sum_{i=1}^N \left(-\frac{\vect{p_{i}}}{m}\cdot\grad_{\vect{q}_i}\phi + m^2\frac{\partial_{\varepsilon_i}[P_i\phi]}{\rho_i^2}\mathfrak{F}_i \right.\\
    &\left. - m^2\smashoperator[l]{\sum_{j=i+1}^N}\left[\frac{P_i}{\rho_i^2}+\frac{P_j}{\rho_j^2}\right]F_{ij}\vect{r}_{ij}\cdot(\grad_{\vect{p}_i}-\grad_{\vect{p}_j})\phi\right).
  \end{aligned}
\]
For $\phi=f_{\mu}$, we have $\partial_{\varepsilon_i}(P_if_{\mu}) = \partial_{\varepsilon_i}(P_i\mathfrak{h})\mathfrak{g} + (P_i\mathfrak{h})(\partial_{\varepsilon_i}\mathfrak{g})$.  
As all the other terms in $\mathcal{L}_{\rm cons}^*$ are first order linear differential operators, it holds
\[
  \mathcal{L}_{\rm cons}^*f_{\mu} = (\mathcal{L}_{\rm cons}^*\mathfrak{h})\mathfrak{g} + \mathfrak{h}(\mathcal{L}_{\rm cons}^*+\mathfrak{L})\mathfrak{g},
\]
where
\[
  \mathfrak{L}\phi = - \sum_{i=1}^N \frac{m^2}{\rho_i^2}\mathfrak{F}_i \partial_{\varepsilon_i}(P_i)\phi ].
\]
We first check that $\mathcal{L}_{\rm cons}^*\mathfrak{h} = 0$.
Since $\mathfrak{h}$ does not depend on $\vect{p}_i$ and thanks to the relations~\eqref{eq:hder-rho2q} and~\eqref{eq:h-derivatives}, we have
\[
\begin{aligned}
  \mathcal{L}_{\rm cons}^* \mathfrak{h} &= -\sum\limits_{i=1}^N\frac{\vect{p_{i}}}{m}\cdot\grad_{\vect{q}_i}\mathfrak{h} + m^2\sum_{i=1}^N\frac1{\rho_i^2}\partial_{\varepsilon_i}(P_i\mathfrak{h})\mathfrak{F}_i\\
  &= \sum\limits_{i=1}^N\left(m\partial_{\rho_i}h + m^2\frac{P_i}{\rho_i^2} \partial_{\varepsilon_i}h + m^2h\frac{\partial_{\varepsilon_i}P_i}{\rho_i^2}\right)\mathfrak{F}_i\\
  &= \sum\limits_{i=1}^N \mathfrak{F}_i\mathcal{A}_{\rm cons}^i h
\end{aligned}
\]
where, using again~\eqref{eq:h-derivatives}, 
\[
\begin{aligned}
  \mathcal{A}_{\rm cons}^i h =& \,m\partial_{\rho_i}h + m^2\frac{P_i}{\rho_i^2} \partial_{\varepsilon_i}h + m^2h\frac{\partial_{\varepsilon_i}P_i}{\rho_i^2}\\
  =& \left[\frac{1}{k_{\rm B}}(\partial_{\rho_i}S_i)(\partial_{\varepsilon_i}S_i) + \partial_{\rho_i}\partial_{\varepsilon_i}S_i +\frac{m}{k_{\rm B}}\frac{P_i}{\rho_i^2}(\partial_{\varepsilon_i}S_i)^2\right.\\
  &\,\left. +\, m\frac{P_i}{\rho_i^2}\partial_{\varepsilon_i}^2S_i +\frac{m}{\rho_i^2}(\partial_{\varepsilon_i}P_i)(\partial_{\varepsilon_i}S_i)\right] \frac{mh}{\partial_{\varepsilon_i}S_i}\\
  =& \left[ \left(\partial_{\rho_i}S_i + m\frac{P_i}{\rho_i^2}\partial_{\varepsilon_i}S_i\right)\frac{\partial_{\varepsilon_i}S_i}{k_{\rm B}} + \partial_{\rho_i}\partial_{\varepsilon_i}S_i\right.\\
  &\,\left.+\, m\frac{P_i}{\rho_i^2}\partial_{\varepsilon_i}^2S_i +\frac{m}{\rho_i^2}(\partial_{\varepsilon_i}P_i)(\partial_{\varepsilon_i}S_i)\right] \frac{mh}{\partial_{\varepsilon_i}S_i}.\\
\end{aligned}
\]
Note that $\mathcal{A}_{\rm cons}^i h$ only involves derivatives of $h$ with respect to the density and energy of particle $i$. We can make use of the relations~\eqref{eq:eos-thermo} to get
\[
\partial_{\rho_i}S_i + m\frac{P_i}{\rho_i^2}\partial_{\varepsilon_i}S_i = 0
\]
and
\[
\begin{aligned} 
  m\frac{\partial_{\varepsilon_i}P_i}{\rho_i^2} &= -\partial_{\varepsilon_i}\left(\frac{\partial_{\rho_i}S_i}{\partial_{\varepsilon_i}S_i}\right)\\
  &= \partial_{\rho_i}S_i\frac{\partial_{\varepsilon_i}^2S_i}{(\partial_{\varepsilon_i}S_i)^2} - \frac{\partial_{\varepsilon_i}\partial_{\rho_i}S_i}{\partial_{\varepsilon_i}S_i}, \\
  &= - m\frac{P_i}{\rho_i^2}\frac{\partial_{\varepsilon_i}^2S_i}{\partial_{\varepsilon_i}S_i} - \frac{\partial_{\varepsilon_i}\partial_{\rho_i}S_i}{\partial_{\varepsilon_i}S_i}.
\end{aligned} 
\]
This shows that $\mathcal{A}_{\rm cons}^i h = 0$.

Now, for a general function $g(E,\vect{P})$, we compute
\[
\begin{aligned}
  &(\mathcal{L}_{\rm cons}^*+\mathfrak{L}) \mathfrak{g} \\
  &\quad = \smashoperator{\sum_{1\leq i<j \leq N}} m^2\left(\frac{P_i}{\rho_i^2}+\frac{P_j}{\rho_j^2}\right)F_{ij}\vect{r}_{ij}\left(\grad_{\ptot}g + \frac{\vect{p}_j}{m}\partial_E g\right)\\
  &\qquad \,- \smashoperator{\sum_{1\leq i<j \leq N}}m^2\left(\frac{P_i}{\rho_i^2}+\frac{P_j}{\rho_j^2}\right)F_{ij}\vect{r}_{ij}\left(\grad_{\ptot}g+ \frac{\vect{p}_i}{m}\partial_E g\right)\\
  &\qquad \,+ \sum_{i=1}^N m^2\frac{P_i}{\rho_i^2}\partial_E g\sum_{j\neq i}F_{ij}\vect{r}_{ij}\cdot\vect{v}_{ij},
\end{aligned}
\]
which clearly vanishes.

The elementary conservative dynamics~\eqref{eq:sdpd-nrj-cons} thus keeps the measure~\eqref{eq:sdpd-energy-minv} invariant.

\subsection{Invariance by the fluctuation / dissipation part of the dynamics}
\label{sec:minv-fd}

We follow here the same proof as for DPDE~\cite{avalos_1997,espanol_1997}.
We rewrite the elementary dynamics~\eqref{eq:sdpd-simple-fluct} using the variable $X_t=(\vect{v}_i^T,\vect{v}_j^T,\varepsilon_i,\varepsilon_j)^T$:
\[
  dX_t = b(X_t)\,\rd t + \mathscr{S}(X_t)\,\rd\vect{W}_t
\]
with
\[
  b(X) = \begin{pmatrix}
    \displaystyle -\mtx{\Gamma}_{ij}\vect{v}_{ij}\\[.5em]
    \displaystyle \mtx{\Gamma}_{ij}\vect{v}_{ij}\\[.5em]
    \displaystyle \frac12\left(\vect{v}_{ij}^T\mtx{\Gamma}_{ij}\vect{v}_{ij} - \frac1{m}\Tr(\mtx{\Sigma}_{ij}\mtx{\Sigma}_{ij}^T)\right)\\[.5em]
    \displaystyle \frac12\left(\vect{v}_{ij}^T\mtx{\Gamma}_{ij}\vect{v}_{ij} - \frac1{m}\Tr(\mtx{\Sigma}_{ij}\mtx{\Sigma}_{ij}^T)\right)
  \end{pmatrix},
\]
and
\[
  \mathscr{S}(X) = \begin{pmatrix}
    \displaystyle \mtx{\Sigma}_{ij}\\[.5em]
    \displaystyle -\mtx{\Sigma}_{ij}\\[.5em]
    \displaystyle -\frac12\vect{v}_{ij}^T\mtx{\Sigma}_{ij}\\[.5em]
    \displaystyle -\frac12\vect{v}_{ij}^T\mtx{\Sigma}_{ij}
  \end{pmatrix}.
\]

The generator for the dynamics~\eqref{eq:sdpd-simple-fluct} is then given by
\[
\mathcal{L}_{{\rm fd},ij} = b\cdot\nabla_X + \frac12\mathscr{S}\mathscr{S}^T:\nabla_X^2,
\]
the matrix $\mathscr{S}\mathscr{S}^T$ being
\[
  \begin{aligned}
    &\mathscr{S}\mathscr{S}^T =\\
    &\quad\begin{pmatrix}
      \displaystyle \mtx{\Sigma}\mtx{\Sigma}^T & \displaystyle -\mtx{\Sigma}\mtx{\Sigma}^T & \displaystyle -\frac12\mtx{\Sigma}\mtx{\Sigma}^T\vect{v} & \displaystyle \frac12\mtx{\Sigma}\mtx{\Sigma}^T\vect{v}\\[.5em]
      \displaystyle  -\mtx{\Sigma}\mtx{\Sigma}^T & \displaystyle \mtx{\Sigma}\mtx{\Sigma}^T & \displaystyle \frac12\mtx{\Sigma}\mtx{\Sigma}^T\vect{v} &\displaystyle  -\frac12\mtx{\Sigma}\mtx{\Sigma}^T\vect{v}\\[.5em]
      \displaystyle -\frac12\vect{v}^T\mtx{\Sigma}\mtx{\Sigma}^T & \displaystyle \frac12\vect{v}^T\mtx{\Sigma}\mtx{\Sigma}^T & \displaystyle \frac14\vect{v}^T\mtx{\Sigma}\mtx{\Sigma}^T\vect{v} & \displaystyle \frac14\vect{v}^T\mtx{\Sigma}\mtx{\Sigma}^T\vect{v}\\[.5em]
      \displaystyle \frac12\vect{v}^T\mtx{\Sigma}\mtx{\Sigma}^T & \displaystyle -\frac12\vect{v}^T\mtx{\Sigma}\mtx{\Sigma}^T & \displaystyle \frac14\vect{v}^T\mtx{\Sigma}\mtx{\Sigma}^T\vect{v} & \displaystyle \frac14\vect{v}^T\mtx{\Sigma}\mtx{\Sigma}^T\vect{v}
    \end{pmatrix},
  \end{aligned}
\]
where we omitted the indices $(i,j)$.
More explicitly, assuming $\mtx{\Gamma}_{ij}$ to be symmetric,
\[
\begin{aligned}
  \mathcal{L}_{{\rm fd},ij} =& -\vect{v}_{ij}\mtx{\Gamma}_{ij}(\grad_{\vect{p}_i}-\grad_{\vect{p}_j})\\
  &+ \frac12\left[\vect{v}_{ij}^T\mtx{\Gamma}_{ij}\vect{v}_{ij} -\frac1{m}{\rm Tr}\left(\mtx{\Sigma}_{ij}\mtx{\Sigma}_{ij}^T\right)\right](\partial_{\varepsilon_i}+\partial_{\varepsilon_j})\\
  &+\frac12(\grad_{\vect{p}_i}-\grad_{\vect{p}_j})^T\mtx{\Sigma}_{ij}\mtx{\Sigma}_{ij}^T(\grad_{\vect{p}_i}-\grad_{\vect{p}_j})\\
  &- \frac12 \vect{v}_{ij}^T\mtx{\Sigma}_{ij}\mtx{\Sigma}_{ij}^T(\grad_{\vect{p}_i}-\grad_{\vect{p}_j})(\partial_{\varepsilon_i}+\partial_{\varepsilon_j})\\
  &+\frac18 \vect{v}_{ij}^T\mtx{\Sigma}_{ij}\mtx{\Sigma}_{ij}^T \vect{v}_{ij} (\partial_{\varepsilon_i}+\partial_{\varepsilon_j})^2.
\end{aligned}
\]
Introducing
\[
\vect{A}_{ij} = \grad_{\vect{p}_i}-\grad_{\vect{p}_j} - \frac12 \vect{v}_{ij} (\partial_{\varepsilon_i}+\partial_{\varepsilon_j}),
\]
the generator can be rewritten as
\[
  \mathcal{L}_{{\rm fd},ij} = -\vect{v}_{ij}^T\mtx{\Gamma}_{ij}\vect{A}_{ij} + \frac12\mtx{\Sigma}_{ij}\mtx{\Sigma}_{ij}^T:\vect{A}_{ij}\vect{A}_{ij}^T.
\]

We now consider matrices of the form~\eqref{eq:fluct-gamma} and use the following relations
\[
  \mtx{\Gamma}_{ij} = \gamma^{\parallel}_{ij}\pj^{\parallel}_{ij} + \gamma^{\perp}_{ij}\pj^{\perp}_{ij},\quad 
  \mtx{\Sigma}_{ij}\mtx{\Sigma}_{ij}^T = (\sigma^{\parallel}_{ij})^2\pj^{\parallel}_{ij} + (\sigma^{\perp}_{ij})^2\pj^{\perp}_{ij},
\]
to write the generator $\mathcal{L}_{{\rm fd},ij}$ as a sum of two operators $\mathcal{L}_{\perp,ij}$ and $\mathcal{L}_{\parallel,ij}$, more precisely $\mathcal{L}_{{\rm fd},ij} = \mathcal{L}_{\perp,ij} + \mathcal{L}_{\parallel,ij}$ where
\[
  \mathcal{L}_{\theta,ij} = -\gamma_{ij}^{\theta}\vect{v}_{ij}^T\pj_{ij}^{\theta}\vect{A}_{ij} + \frac12\left(\sigma_{ij}^{\theta}\right)^2\vect{A}_{ij}^T\pj_{ij}^{\theta}\vect{A}_{ij},
\]
for $\theta \in \{ \perp, \parallel \}$.
Since $\vect{A}_{ij}^* = -\vect{A}_{ij}$, the adjoint of the generator applied to $f_{\mu}$ reads
\[
  \mathcal{L}_{\theta,ij}^* f_{\mu} = \vect{A}_{ij}^T\pj_{ij}^{\theta} \left(\gamma_{ij}^{\theta}f_{\mu}\vect{v}_{ij} + \frac12\vect{A}_{ij}\left[\left(\sigma_{ij}^{\theta}\right)^2f_{\mu}\right]\right).
\]

A sufficient condition for the measure $\mu$ to be left invariant by~\eqref{eq:sdpd-simple-fluct} is then
\[
  \gamma^{\theta}_{ij}\vect{v}_{ij}f_{\mu} + \frac12 \vect{A}_{ij}[(\sigma^{\theta}_{ij})^2f_{\mu}] = 0
\]
for $\theta \in \{\perp, \parallel\}$.
Since $\vect{A}_{ij}$ is a first order linear differential operator, we can write
\[
  \begin{aligned}
    \vect{A}_{ij}\left[ \left(\sigma_{ij}^{\theta}\right)^2f_{\mu} \right] &=  \vect{A}_{ij}\left[ \left(\sigma_{ij}^{\theta}\right)^2\mathfrak{gh} \right]\\
    &= \left(\vect{A}_{ij}\left[\left(\sigma_{ij}^{\theta}\right)^2\mathfrak{h}\right]\right)\mathfrak{g} + \left(\sigma_{ij}^{\theta}\right)^2\mathfrak{h} \left(\vect{A}_{ij}\mathfrak{g}\right).
  \end{aligned}
\]
For any function $\mathfrak{g}$, it holds
\[
  \vect{A}_{ij}\mathfrak{g} = \left(\frac{\vect{p}_i}{m}+\frac{\vect{p}_j}{m}\right)\partial_{\etot}g - \vect{v}_{ij} \partial_{\etot}g = \,\vect{0}.
\]
We assume that the fluctuation amplitude $\sigma_{ij}^{\theta}$ depends only on the positions $\vect{q}$ and the internal energies $\varepsilon_i$ and $\varepsilon_j$.
We next compute
\[
\begin{aligned}
  \vect{A}_{ij}\left[(\sigma_{ij}^{\theta})^2\mathfrak{h}\right] =&\, \vect{A}_{ij}\left((\sigma_{ij}^{\theta})^2\prod_{k=1}^N \frac{\exp\left(\frac{S_k(\varepsilon_k,\vect{q})}{k_{\rm B}}\right)}{T_k(\varepsilon_k,\vect{q})}\right)\\
  =& -\frac12\mathfrak{h}\vect{v}_{ij} \left(T_iT_j(\partial_{\varepsilon_i}+\partial_{\varepsilon_j})\left[\frac{(\sigma_{ij}^{\theta})^2}{T_iT_j}\right]\right.\\
  &\qquad\qquad\left. + (\sigma_{ij}^{\theta})^2\left[ \frac{\partial_{\varepsilon_i}S_i}{k_{\rm B}} + \frac{\partial_{\varepsilon_j}S_j}{k_{\rm B}} \right] \vphantom{\left[\frac{(\sigma_{ij}^{\theta})^2}{T_iT_j}\right]}\right)\\
  =& -\frac12 \mathfrak{h}\vect{v}_{ij}\left( T_iT_j(\partial_{\varepsilon_i}+\partial_{\varepsilon_j})\left[\frac{(\sigma_{ij}^{\theta})^2}{T_iT_j}\right]\right.\\
  &\qquad\qquad\left. +  \frac{(\sigma_{ij}^{\theta})^2}{k_{\rm B}} \frac{T_i+T_j}{T_iT_j} \right)
\end{aligned}
\]
This leads us to the following  sufficient condition on $\gamma_{ij}^{\theta}$ and $\sigma_{ij}^{\theta}$:
\begin{equation}
  \label{eq:generic-gamma}
  \gamma_{ij}^{\theta} = \frac14 \left(T_iT_j(\partial_{\varepsilon_i}+\partial_{\varepsilon_j})\left[\frac{(\sigma_{ij}^{\theta})^2}{T_iT_j}\right] +  \frac{(\sigma_{ij}^{\theta})^2}{k_{\rm B}} \frac{T_i+T_j}{T_iT_j} \right).
\end{equation}
There are many possible solutions to this equation.
Actually any choice for $\sigma_{ij}^{\theta}$ (for example $\sigma_{ij}^{\theta}$ constant as in DPDE) yields a corresponding expression for $\gamma_{ij}^{\theta}$.

The expression of the fluctuation amplitude in the original SDPD~\eqref{eq:sdpd-fluct-coefficient} suggests taking
\[
\sigma^{\theta} = \sqrt{\kappa^{\theta}_{ij} k_{\rm B}\frac{T_iT_j}{T_i+T_j}},
\]
with $\kappa_{ij}^{\theta}$ a positive constant.
We can then further evaluate
\[
  \begin{aligned}
    \gamma_{ij}^{\theta} &= \frac14\left( T_iT_j(\partial_{\varepsilon_i}+\partial_{\varepsilon_j})\left[\frac{\kappa^{\theta}_{ij} k_{\rm B}}{T_i+T_j}\right] + \kappa^{\theta}_{ij} \right)\\
    &= \frac14\kappa_{ij}^{\theta}\left( 1 - k_{\rm B}\left(\frac1{C_i} + \frac1{C_j}\right)\frac{T_iT_j}{(T_i+T_j)^2}\right),
  \end{aligned}
\]
by using the expressions of the heat capacities~\eqref{eq:eos-thermo}.
More precisely,
\[
  \partial_{\varepsilon_i}\left(\frac1{T_i+T_j}\right) = - \frac{\partial_{\varepsilon_i}T_i}{(T_i+T_j)^2} = -\frac1{C_i}\frac1{(T_i+T_j)^2}.
\]
In order to retrieve the friction term of the original SDPD dynamics~\eqref{eq:sdpd-espanol}, we choose
\[
  \kappa^{\perp}_{ij} = 4a_{ij}, \quad \text{and} \quad
  \kappa^{\parallel}_{ij} = \frac43a_{ij}+b_{ij}.
\]
This choice for the friction and fluctuation coefficients ensures that each pairwise elementary dynamics leaves the measure~\eqref{eq:sdpd-energy-minv} invariant.

\section{Evaluation of thermodynamic properties}
\label{sec:app-tp}
We gather in this appendix the computation of estimators for thermodynamic quantities.
We focus on the internal temperature in Section~\ref{sec:tp-temp} and on pressure in Section~\ref{sec:tp-pressure}.
In the following, $\mathscr{E} = \Omega^N\times \mathbb{R}^{3N}\times \mathbb{R}_+^N$ stands for the phase space as defined in Section~\ref{sec:nrj-sdpd}.

\subsection{Internal Temperature}
\label{sec:tp-temp}
The ensemble average of the internal temperature under the canonical measure $\mu_{\beta}$ defined in~\eqref{eq:sdpd-minveq} is given by
\begin{widetext}
\begin{equation}
  \label{eq:sdpd-temperature}
  \begin{aligned}
    \langle T_i \rangle_{\mu_{\beta}} &= Z_{\beta}^{-1}\displaystyle \int_{\mathscr{E}} T_i(\varepsilon_i,\vect{q}) \exp\left(-\beta \sum_{j=1}^N\frac{\vect{p}_j^2}{m}\right) \prod_{j=1}^N\frac{\exp\left(\frac{S_j(\varepsilon_j,\vect{q})}{k_{\rm B}}-\beta \varepsilon_j\right)}{T_j(\varepsilon_j,\vect{q})}\,\rd\vect{q}\,\rd\vect{p}\,\rd\varepsilon,\\
    &= \mathfrak{Z}_{\beta}^{-1} \int_{\vect{q}\in\Omega^N} \left[\int_{\mathbb{R}_+}\exp\left(\frac{S_i(\varepsilon_i,\vect{q})}{k_{\rm B}}-\beta \varepsilon_i\right)\rd\varepsilon_i\right] \prod\limits_{j\neq i}\mathcal{R}_j(\vect{q})\,\rd\vect{q},
  \end{aligned}
\end{equation}
\end{widetext}
where we have integrated out the momenta and introduced
\[
\mathcal{R}_j(\vect{q}) = \int_{\mathbb{R}_+}\frac{\exp\left(\frac{S_j(\varepsilon_j,\vect{q})}{k_{\rm B}}-\beta \varepsilon_j\right)}{T_j(\varepsilon_j,\vect{q})}\,\rd\varepsilon_j,
\]
along with the normalization constant $\mathfrak{Z}_{\beta} = \int_{\Omega^N}\prod_{j=1}^N\mathcal{R}_j(\vect{q})\,\rd\vect{q}$.
A simple computation based on~\eqref{eq:eos-thermo} gives
\[
\begin{aligned}
  \mathcal{R}_j(\vect{q}) =& \int_0^{+\infty}\exp\left(\frac{S_j(\varepsilon_j,\vect{q})}{k_{\rm B}}-\beta \varepsilon_j\right)(\partial_{\varepsilon_j}S_j)\,\rd\varepsilon_j,\\
  =&\, k_{\rm B}\left( \int_0^{+\infty}\partial_{\varepsilon_j}\left[\exp\left(\frac{S_j(\varepsilon_j,\vect{q})}{k_{\rm B}}-\beta \varepsilon_j\right)\right] \rd\varepsilon_j\right.\\
  &\left. + \beta \int_0^{+\infty}\exp\left(\frac{S_j(\varepsilon_j,\vect{q})}{k_{\rm B}}-\beta \varepsilon_j\right)\rd \varepsilon_j\right).
\end{aligned}
\]
Under the assumptions~\eqref{eq:entropy-assumptions}, we therefore obtain
\begin{equation}
  \label{eq:rj}
  \mathcal{R}_j(\vect{q}) = k_{\rm B}\beta\int_0^{+\infty}\exp\left(\frac{S_j(\varepsilon_j,\vect{q})}{k_{\rm B}}-\beta \varepsilon_j\right)\,\rd \varepsilon_j.
\end{equation}
Plugging this result in~\eqref{eq:sdpd-temperature} leads to
\[
\langle T_i \rangle_{\mu_{\beta}} = \frac1{k_{\rm B}\beta} = T_{\beta}.
\]

\subsection{Pressure}
\label{sec:tp-pressure}
To evaluate the pressure in the SDPD system, we first compute the partition function of the canonical measure $\mu_{\beta}$~\eqref{eq:sdpd-minveq}:
\[
\begin{aligned}
  Z_{\beta} =& \int_{\mathscr{E}} \frac{\displaystyle \exp\left(\sum_{i=1}^N -\beta\left[\frac{\vect{p}_i^2}{2m} + \varepsilon_i\right] + \frac{S_i(\varepsilon_i,\vect{q})}{k_{\rm B}}\right)}{\displaystyle \prod\limits_{i=1}^NT_i(\varepsilon_i,\vect{q})}\,\rd\vect{q}\,\rd\vect{p}\,\rd\varepsilon\\
  =& \int_{\mathbb{R}^{3N}}\exp\left(-\beta\sum_{i=1}^N \frac{\vect{p}_i^2}{2m}\right)\,\rd\vect{p}\\
  &\times \int_{\Omega^N} \prod_{i=1}^N \left[\int_{\mathbb{R}_+} \frac {\exp\left(-\beta \varepsilon_i + \frac{S_i(\varepsilon_i,\vect{q})}{k_{\rm B}}\right)} {T_i(\varepsilon_i,\vect{q})}\,\rd\varepsilon_i\right]\rd\vect{q}.
\end{aligned}
\]

From the results of Section~\ref{sec:tp-temp}  (see~\eqref{eq:rj}),
\begin{equation}
  \label{eq:partition-energy-intg}
  \begin{aligned}
    &\int_{\mathbb{R}_+}\frac{\exp\left(-\beta \varepsilon_i + \frac{S_i(\varepsilon_i,\vect{q})}{k_{\rm B}}\right)}{T_i(\varepsilon_i,\vect{q})}\,\rd \varepsilon_i\\
    &\qquad = \frac1{T_{\beta}}\int_{\mathbb{R}_+}\exp\left(-\beta \varepsilon_i + \frac{S_i(\varepsilon_i,\vect{q})}{k_{\rm B}}\right)\rd\varepsilon_i,
  \end{aligned}
\end{equation}
which allows us to rewrite the partition function as
\[
\begin{aligned}
  Z_{\beta} &= \mathcal{Z}_{\beta}\int_{\Omega^N}\prod_{i=1}^N\left[\int_{\mathbb{R}_+}\exp\left(-\beta \varepsilon_i + \frac{S_i(\varepsilon_i,\vect{q})}{k_{\rm B}}\right)\rd\varepsilon_i\right]\rd\vect{q},
\end{aligned}
\]
where $\displaystyle \mathcal{Z}_{\beta} = \frac1{T_{\beta}^{N}}\left(2\pi\frac{m}{\beta}\right)^{-3N/2}$ does not depend on the volume $\mathcal{V}= \abs{\Omega}$ of the domain.

The free energy is given by
\[
\mathcal{F}_{\beta}(\beta,\mathcal{V}) = -\frac1{\beta}{\rm log}Z_{\beta}(\beta,\mathcal{V})
\]
and the thermodynamic pressure by
\begin{equation}
  \label{eq:thermo-pressure}
    \mathcal{P}_{\beta} = -\partial_{\mathcal{V}}\mathcal{F}_{\beta} = \frac1{\beta}\frac{\partial_{\mathcal{V}}Z_{\beta}}{Z_{\beta}}.
\end{equation}

In order to compute the derivative of $Z_{\beta}$ with respect to the volume $\mathcal{V}$ of the system, we introduce a spatial dilation $\vect{\tilde{q}}=(1+\lambda)\vect{q}$ for $\lambda > -1$.
We then consider the partition function associated with the domain $(1+\lambda)\Omega$:
\[
\mathscr{Z}(\lambda) = \mathcal{Z}_{\beta}  \smashoperator[l]{\int\limits_{[(1+\lambda)\Omega]^N}}\hspace{-.3em}\prod_{i=1}^N \left[\int_{\mathbb{R}_+}\hspace{-.5em}\exp\left(\frac{S_i(\varepsilon_i,\vect{\tilde{q}})}{k_{\rm B}}-\beta \varepsilon_i \right)\,\rd \varepsilon_i\right]\rd\vect{\tilde{q}}
\]
which gives, after a change of variables allowing to map back the integration domain to $\Omega^N$:
\[
  \begin{aligned}
    &\mathscr{Z}(\lambda) = \mathcal{Z}_{\beta} (1+\lambda)^{3N}\times\\
    &\quad \int_{\Omega^N} \prod_{i=1}^N \left[\int_{\mathbb{R}_+}\exp\left(-\beta \varepsilon_i + \frac{S_i(\varepsilon_i,(1+\lambda)\vect{q})}{k_{\rm B}}\right)\,\rd \varepsilon_i\right]\,\rd\vect{q}.
  \end{aligned}
\]
Note that $\lambda=0$ corresponds to no dilation and thus $\mathscr{Z}(0) = Z_{\beta}$.
Since the volume of the dilated domain is given by
\[
  \mathcal{V}(\lambda) = (1+\lambda)^3\mathcal{V}(0),
\]
it holds $\mathcal{V}'(0) = 3\mathcal{V}(0)$.
The derivative of the partition function $Z_{\beta}$ with respect to the volume $\mathcal{V}$ can then be written as
\[
  \partial_{\mathcal{V}}Z_{\beta} = \frac{\mathscr{Z}'(0)}{\mathcal{V}'(0)} = \frac{\mathscr{Z}'(0)}{3\mathcal{V}(0)}.
\]
We can now derive ${\rm log}(\mathscr{Z}(\lambda))$ with respect to $\lambda$ at $\lambda=0$ as
\begin{equation}
  \label{eq:zlambda-deriv}
  \begin{aligned}
    \frac{\mathscr{Z}'(0)}{\mathscr{Z}(0)} =&\, 3N + \frac{\mathcal{Z}_{\beta}}{ Z_{\beta}} \hspace{-.2em} \int\limits_{\Omega^N\times\mathbb{R}_+^N} \hspace{-.8em} \left.\partial_{\lambda}\left(\sum_{i=1}^N \frac{S_i(\varepsilon_i,(1+\lambda)\vect{q})}{k_{\rm B}}\right)\right|_{\lambda=0}\\
    &\,\times \exp\left(\sum\limits_{i=1}^N -\beta \varepsilon_i + \frac{S_i(\varepsilon_i,\vect{q})}{k_{\rm B}}\right)\rd\varepsilon\,\rd\vect{q}.
  \end{aligned}
\end{equation}
In order to evaluate~\eqref{eq:zlambda-deriv}, we first need to compute the derivatives of the density $\rho_i$ and of the entropy $S_i$ with respect to $\lambda$.
We have, by deriving equation~(\ref{eq:sdpd-rho-v}),
\[
\begin{aligned}
  \partial_{\lambda}\left[\rho_i\left((1+\lambda)\vect{q}\right)\right] &= \sum_{j\neq i} m \partial_{\lambda}[ W((1+\lambda)\vect{r}_{ij}) ]\\
  &= -(1+\lambda)\sum_{j\neq i}mF((1+\lambda)\vect{r}_{ij})\vect{r}_{ij}^2.
\end{aligned}
\]
With this result and the equation of state~\eqref{eq:sdpd-eos}, we can compute $\partial_{\lambda}S_i(\varepsilon_i,(1+\lambda)\vect{q})$ as
\[
\begin{aligned}
  \partial_{\lambda} [ S_i\left(\varepsilon_i,(1+\lambda)\vect{q}\right) ] &= \partial_{\lambda}[\rho_i((1+\lambda)\vect{q})]\partial_{\rho_i}S_i\\
  &= (1+\lambda)\sum_{j\neq i}\frac{m^2P_i}{\rho_i^2T_i}F((1+\lambda)\vect{r}_{ij})\vect{r}_{ij}^2.
\end{aligned}
\]
Finally, plugging these expressions evaluated at $\lambda=0$ in equation~\eqref{eq:zlambda-deriv} leads to
\begin{widetext}
  \[
    \begin{aligned}
      \frac{\mathscr{Z}'(0)}{\mathscr{Z}(0)} =&\, 3N + Z_{\beta}^{-1} \frac1{k_{\rm B}T_{\beta}^{N}}\left(2\pi\frac{m}{\beta}\right)^{-3N/2} \left(\int_{\Omega^N}\int_{\mathbb{R}_+^N} \left[ \sum\limits_{i=1}^N\sum\limits_{j\neq i}m^2F_{ij}\vect{r}_{ij}^2\frac{P_i}{\rho_i^2T_i} \right] \prod\limits_{k=1}^N \exp\left( -\beta \varepsilon_k + \frac{S_k(\varepsilon_k,\vect{q})}{k_{\rm B}}\right)\,\rd\varepsilon\,\rd\vect{q}\right)\\
      =&\, 3N + \sum_{i=1}^N \beta Z_{\beta}^{-1}\int_{\mathscr{E}} \left[\frac{P_i}{\rho_i^2}\sum_{j\neq i}m^2F_{ij}\vect{r}_{ij}^2\right] \exp\left(-\beta\sum_{k=1}^N \left[\frac{\vect{p}_k^2}{2m} + \varepsilon_k\right]\right) \frac{\exp\left(\frac{S_i}{k_{\rm B}}\right)}{T_i} \prod_{k\neq i} \frac{\exp\left(\frac{S_k}{k_{\rm B}}\right)}{T_{\beta}} \,\rd\vect{q}\,\rd\vect{p}\,\rd \varepsilon.
    \end{aligned}
  \]
\end{widetext}
In each term of the sum, we use \eqref{eq:partition-energy-intg} for all variables $\varepsilon_k$ with $k\neq i$, which gives
\[
\frac{\mathscr{Z}'(0)}{\mathscr{Z}(0)} = 3N + \beta\int_{\mathscr{E}} \left[\sum_{i=1}^N\sum_{j\neq i}m^2F_{ij}\vect{r}_{ij}^2\frac{P_i}{\rho_i^2}\right]\mu_{\beta}(\rd\vect{q}\,\rd\vect{p}\,\rd\varepsilon).
\]

We can finally compute the thermodynamic pressure in a SDPD system using~\eqref{eq:thermo-pressure}:
\[
  \mathcal{P} = \frac{1}{3\beta\mathcal{V}}\frac{\mathscr{Z}'(0)}{\mathscr{Z}(0)} = \langle P_{\rm kin}\rangle_{\mu_{\beta}} +  \langle P_{\rm virial}\rangle_{\mu_{\beta}},
\]
where we identify the usual expressions for the kinetic part of the pressure:
\[
  \langle P_{\rm kin}\rangle_{\mu_{\beta}} = \frac{N}{\mathcal{\beta\mathcal{V}}},
\]
and for the potential part of the pressure:
\[
  \begin{aligned}
    P_{\rm virial} &= \frac1{3\mathcal{V}}\sum_{i=1}^Nm^2\frac{P_i}{\rho_i^2}\sum_{j\neq i}F_{ij}\vect{r}_{ij}^2\\
    &= \frac1{3\mathcal{V}}\sum\limits_{1\leq i<j\leq N}\vect{\mathcal{F}}_{{\rm cons},ij}\cdot\vect{r}_{ij},
  \end{aligned}
\]
with the forces $\vect{\mathcal{F}}_{{\rm cons},ij}$ defined in~(\ref{eq:cons-forces}).

\medskip

%

\end{document}